\documentstyle[preprint,aps,floats,epsfig,epsf]{revtex}
\tightenlines
\begin{document}
\draft

\preprint{\vbox{\hbox{JLAB-THY-02-01}}}

\title{Quark-Hadron Duality in a Relativistic,
Conf\/ining Model}

\author{Sabine Jeschonnek$^{(1)}$
 and J. W. Van Orden$^{(2,3)}$}

\address{\small \sl
(1) The Ohio State University, Physics Department, Lima OH 45804 \\
(2) Jefferson Lab, 12000 Jefferson Ave, Newport News, VA 23606 \\
(3) Department of Physics, Old Dominion University, Norfolk, VA 23529}

\date{\today}
\maketitle

\begin{abstract}

{\it This paper is dedicated to the memory of our
collaborator Nathan Isgur.}
\newline

Quark-hadron duality is an interesting and potentially very useful
phenomenon, as it relates the properly averaged hadronic data to
a perturbative QCD result in some kinematic regimes.
While duality is well
established experimentally, our current theoretical understanding is still
incomplete. We employ a simple model to qualitatively reproduce all
the features of Bloom-Gilman duality as seen in electron scattering.
In particular, we address the role of relativity, give an explicit
analytic proof of the equality of the hadronic and partonic scaling curves,
and show how the transition from coherent to incoherent scattering
takes place.

\end{abstract}

\section{Introduction}

Quark-hadron duality has been well established experimentally
\cite{bgduality} for over 30 years, but our theoretical understanding
of the phenomenon is quite limited so far. In the past year, there has
been renewed interest in duality, both on the experimental
\cite{jlab,newexp} and theoretical side
\cite{dp1,closeisgur,pp,simula,wallym}. Duality is a
major point in the planned 12 GeV upgrade of CEBAF at Jefferson Lab
\cite{12gevwp}.
Duality between partons and hadrons is also employed in QCD sum rules.
\cite{qcdsr}.

In a recent publication \cite{dp1}, we presented results obtained in a
confining, relativistic model which qualitatively reproduced the
features seen in Bloom-Gilman duality.
In \cite{dp1}, we only discussed a reaction where all particles
involved - electrons, photons, and quarks - were treated as
scalars. In this paper, we present the model for physical electrons and
photons, and only treat the quarks as scalars. We describe the model and
its properties in more detail, and discuss the Coulomb sum rule, the
transition from coherent to incoherent scattering, and duality in the
form factors. We also put special emphasis on the role of
relativity. Relativistic treatment was one of four basic conditions
imposed in \cite{dp1} to obtain duality, and here we show the
consequences of relaxing this condition, and compare non-relativistic
and relativistic calculations. One main point of this paper is the
explicit derivation of the scaling curve for scattering from quarks confined
in their initial and final state, and the comparison of this scaling curve to
that obtained in a parton model calculation.

 For the convenience of
the reader, we define the concept of duality in the following, and
briefly discuss a few basic implications.
In the literature, there exist many slightly varying ``definitions''
and usages of the term duality, and the phenomenon manifests itself
experimentally in many different processes. We begin with a
definition that covers all these cases. First, we need to make an
obvious observation: any hadronic process can be correctly described
in terms of quarks and gluons. In other words, quantum chromodynamics
(QCD) is the correct theory for strong interactions. While this
statement is obvious, it has little practical value, as in most cases,
we cannot perform a full QCD calculation. E.g., in order to calculate
a resonance excitation form factor, one would need to include very
many quarks and gluons, and they would all couple strongly.  We will
refer to the above statement that any hadronic process can be
described by a full QCD calculation as ``degrees of freedom duality''.

There exists a more practical and less obvious version of the first
statement: in certain kinematic regions, the average of hadronic
observables is described by a perturbative QCD (pQCD) result.
This is the statement of duality, and we are going to explain the
details in the following.

With pQCD result, we indicate the result for the
underlying quark process -- for inclusive inelastic electron
scattering from a proton, it is free electron-quark scattering; for
semileptonic decays, e.g. $\bar{B} \rightarrow X_c l \bar \nu_l$, it
is the underlying quark decay rate, in this case obtained from the
process $b \rightarrow c l \bar \nu_l$ \cite{isgurwise};
for $e^+ e^- \rightarrow hadrons$,
it is the underlying $e^+ e^- \rightarrow q \bar{q}$ process.
  Now, it is clear that we
expect perturbative QCD to describe Nature in a certain kinematic
regime, i.e. for very large $Q^2$. In this regime, due to
the fact that full QCD is approximated by perturbative QCD, the
statement of duality turns into the statement of the ``degrees of
freedom duality''. So we have identified one kinematic regime in which
even the non-obvious version of duality must hold.

We also can identify a kinematic regime for which duality cannot
hold: for $Q^2 \rightarrow 0$.
While the underlying reason for the breakdown of duality at low four-momentum
transfers is the non-perturbative, strong interaction of the hadrons, one
can see the breakdown of duality most easily by considering the transition
from incoherent to coherent scattering.
For duality to hold for the
nucleon structure functions in this case, we would need the following:
the elastic proton and neutron form factors, which take the value of
the nucleon charge for $Q^2 \rightarrow 0$, would have to be
reproduced by electron scattering off the corresponding $u$ and $d$
quarks. Now, for the proton this can work, as the squares of the
charges of two $u$ quarks and one $d$ quark add up to 1. However, for
the neutron, the squared quark charges cannot add up to 0, so it is
clear that duality in inclusive inelastic electron scattering from a
neutron must fail for $Q^2 \rightarrow 0$.
In addition, we know from gauge invariance that for $Q^2 \rightarrow
0$, at fixed energy transfer $\nu$, the function $\nu W_2 (\nu, Q^2)$
must approach 0. It is clear that the scaling function $F_2 (x)$ does
not show that behavior, which gives us an additional reason to expect
the breakdown of duality at low $Q^2$.

So now we know that
duality has to hold in one kinematic regime and that it has to break
down in another kinematic regime. Obviously, a very interesting
question is what happens in between these regimes, i.e.  how exactly
does duality break down, how far does it hold in the regime where it
is nontrivial, i.e. for moderate values of $Q^2$, and how accurately
does it hold where it holds.

\section{The Model}
\label{secmod}

Here, we present a model for the study of quark-hadron duality that
uses only a few basic assumptions. Namely, we assume that it is
necessary to include confinement and relativity in our model, that
it is sufficient to base our model solely on valence quarks, and that
these quarks can be treated as scalars. A model with these features
will not give a realistic description of any data, but it should allow
us to obtain duality and study the critical questions of when and how
accurately duality holds.

Although it is our aim to study duality in electron scattering from
the nucleon, i.e. from a three-quark-system, as a first step we
simplify the problem at hand by substituting two quarks by an
antiquark,
as the representation $ 3 \otimes 3$ in SU(3)
contains the representation $\bar{3}$.
This means
we have a two-body problem now, and we have to solve the
Bethe-Salpeter equation. In the special case of the mass of the antiquark,
$M$, going to infinity, the problem further simplifies to a one-body
problem. In the case of scalar quarks considered here, we obtain a
Klein-Gordon equation. In contrast to \cite{dp1} where we assumed that
all particles involved - electrons, photons and quarks - are scalars,
here we treat only the valence quarks as scalars.
Note that the experiment which our model resembles most would be electron
scattering from a B meson.
Still, we expect to gain valuable insights from
considering this case, and would like to stress that none of the assumptions
we made here prevent us from extending our model to describe more realistic
circumstances.

We have chosen to implement confinement by using a linear
potential, which leads to a relativistic harmonic oscillator solution.
This has the advantage that analytic solutions can be readily obtained
and that a comparison with the non-relativistic case is easily feasible.

We have to solve the Klein-Gordon equation with a scalar potential:
\begin{equation}
\left( \displaystyle{\frac{\partial^2}{\partial t^2}} - \vec
\nabla ^2 + m^2 + V^2 \right ) \Phi(x) = 0\label{KGeqn}
\end{equation}
with the usual ansatz $\Phi^\pm(x) = \Phi(\vec r \, ) \exp(\mp i E
t)$ and the confining potential
\begin{equation}
V^2 (\vec r \,) = b^2 \, r^2,
\end{equation}
where $b$ is the  relativistic string tension and
has dimension $[b] = [\mbox{mass}^2]$. The superscript of the wave
function denotes positive and negative energy solutions to the
Klein-Gordon equation. The mass of the quark is denoted by $m$
and we use $m = 0.33 $ GeV throughout this paper.

The Klein-Gordon equation in this form can be easily rearranged to
have the form of a Schr\"odinger equation to give
\begin{equation}
\left ( - \displaystyle{\frac{\vec \nabla ^2}{2 m}} + \frac{1}{2}
\frac{b^2}{m} r^2 \right)  \, \Phi(\vec r \,) = \frac{E^2-m^2}{2m}
\, \Phi(\vec r \,)\, ,
\end{equation}
where the similarity to the Schr\"odinger equation for a
non-relativistic harmonic oscillator potential becomes apparent.
The solutions to this equation are easily obtained by making the
substitutions $\widetilde \kappa \equiv \frac{b^2}{m}$ and $\widetilde E
\equiv \frac{E^2-m^2}{2m}$ and using the well known solutions of
the non-relativistic case.

The energy eigenvalues for the Klein-Gordon equation are $E=\pm
E_N$ where
\begin{equation}
 E_N = \sqrt{ 2 \beta^2(N+ \frac{3}{2}) + m^2 } \,.
\end{equation}
$N$ is the principal oscillator quantum number and $\beta\equiv
b^{1/2}$. The corresponding wave functions are the usual
nonrelativistic oscillator wave functions. For the present
application, it is convenient to express the oscillator wave
function in Cartesian form as
\begin{equation}
\Phi_N (\vec r \,) = \phi_{n_x} (x) \, \phi_{n_y} (y) \,
\phi_{n_z} (z)
\end{equation}
where
\begin{equation}
\phi_{n_x} (x) = \displaystyle{\frac{\sqrt{\beta}}{\sqrt{2^{n_x}
n_x! \sqrt{\pi}}}} \, H_{n_x} (\beta x) \exp{(-\frac{1}{2}
\beta^2}x^2) \,
\end{equation}
with similar expressions for y and z coordinates. The $H_n$ are
the Hermite polynomials. Unless noted otherwise, we use $\beta =
0.4 $ GeV, which was chosen to give reasonable values for the mass
splitting and charge radius.

It should be noted that the negative energy solutions are just
that since we are using a one-body wave equation and not a field
theory. Therefore, these are an artifact of the model, but are
necessary to provide a complete set of relativistic states.

As we choose to retain the non-relativistic wave functions,
we differ from the usual relativistic normalization:
using the Klein-Gordon normalization condition for these wave
functions gives
\begin{equation}
i \, \int d^3r\left( {\Phi_N^\pm}^* (x)\partial^0\Phi_N^\pm (x)-
\left(\partial^0{\Phi_N^\pm}^* (x)\right)\Phi_N^\pm (x)\right) =
\pm 2E_N
\end{equation}
This leads to the factor $1/ 4 E_0 E_N$ in the response functions, and to
the energy factor in the current operator defined below. Of course,
we could have used explicitly relativistic normalized wave functions,
which would have led to a different expression for the form factor,
given below in Eq.~(\ref{eqff}).

The basic difference between the relativistic and the
non-relativistic oscillator equation is the difference in the
energy spectrum: while the non-relativistic solutions are
equally spaced, as $E_N^{non-rel.} \propto N$, the relativistic
spectrum goes as $E_N \propto \sqrt{N}$ for large $N$ so the
density of states increases with increasing $N$. We note in
passing that with this relativistic spectrum our simple model
gives rise to linear Regge trajectories\cite{collins} as seen in
Nature.

In the following, we will consider electron scattering from a meson
with an infinitely heavy antiquark.  In contrast to our previous
publication \cite{dp1}, where we treated all particles as scalars, the
electrons in this paper are spin 1/2 fermions, and the virtual
photons have spin 1. Unless otherwise noted, we assume in this paper
that only the light
quark carries a charge, and that the photon therefore couples only to
the light quark, not to the heavy antiquark.

For a photon coupling to the quark in the positive energy harmonic
oscillator ground state and transferring the four-momentum
$q=(\nu, \vec q)$, we have the following current matrix element:
\begin{equation}
j^{\mu} (q) = \frac{i}{4 E_0 E_N} \, \int d^4 x \, \exp(-i q \cdot x) \, \left (
\Phi_{N_f}^* (x) \, \partial^{\mu} \Phi_0 (x) - (\partial^{\mu}
 \Phi_{N_f}^* (x)) \, \Phi_0 (x) \right )
\end{equation}
where $N_f$ can designate either a positive or negative energy
state. Using this definition of the current along with
(\ref{KGeqn}), it can be easily shown that the current is
conserved,
$q_{\mu} j^{\mu} = 0$.

The calculation of the double differential cross section is
straightforward and leads to the Rosenbluth equation
\begin{equation}
\frac{d \sigma} {d E_f d \Omega_f} = \sigma_{\rm{Mott}}  \left[ v_L
R_L(\vec q,\nu) +
 v_t R_T(\vec q,\nu)\right] \,, \label{diffxs}
\end{equation}
where $\sigma_{Mott}$ is the Mott cross section, $v_L$ and $v_T$
are the usual leptonic coefficients
$$ v_L = \frac{Q^4}{\vec q \, ^4} \, \, , v_T = \frac{1}{2}
\frac{Q^2}{\vec q \, ^2} + \tan ^2 \frac{\vartheta_e}{2} \, , $$
$Q^2 \equiv -q^2 = \vec q \, ^2 - \nu^2$, and the longitudinal and
transverse response functions are
\begin{equation}
R_L(\vec q,\nu)= \sum_{N=0}^\infty  \frac{1}{4 \, E_0 \, E_N}
 |F_{0N} (\vec q)|^2\left[ (E_0+ E_N)^2 \delta(\nu + E_0 - E_N)-
 (E_0- E_N)^2 \delta(\nu + E_0 + E_N)\right]\,\label{RL}
\end{equation}
and
\begin{equation}
R_T(\vec q,\nu) = 8 \frac{\alpha}{\vec q \, ^2}\sum_{N=0}^\infty
\frac{1}{4 \, E_0 \, E_N}
  \, N \, |F_{0N} (\vec q)|^2\left[  \delta(\nu + E_0  - E_N)-
  \delta(\nu + E_0  + E_N)\right] \,.
\end{equation}
In these expressions, $F_{0,N}$ stands for the excitation form
factor,
\begin{equation}
F_{0,N} (\vec q \, ^2) = \int d^3 \vec r \, \exp(i \vec q \cdot
\vec r) \, \Phi_{N_f}^* (\vec r) \, \Phi_0 (\vec r) = \int d^3
\vec p \, \Psi^*_{N_f} (\vec p) \, \Psi_0 (\vec p - \vec q) \, ,
\label{defff}
\end{equation}
where $\Psi$ indicates a momentum space wave function.
Making use of the recurrence relations of the Hermite
polynomials, we find an explicit expression for $F_{0,N}$:
\begin{equation}
F_{0,N} (\vec q \, ^2) = \frac{1}{\sqrt{N!}} \, i^N \,
\left ( \frac{|\vec q|}{\sqrt{2} \, \beta} \right ) ^N
\exp (- \frac{\vec q \, ^2}{4 \, \beta^2}) \, .
\label{eqff}
\end{equation}
Note that some care is necessary in writing the expressions for the
responses to properly include the negative energy states. The relative
sign between the positive and negative energy contributions is
associated with the negative norm of the negative energy states.

These expressions for the response functions have been derived
assuming that the quark is excited from the ground state into a
resonance state, $N$, and remains there without decaying. This is
just the first step on the way to meson production in this
picture. The $\delta$-function in the energies is an artifact of
this assumption.

Note that because we assume scalar quarks, there is no magnetization
current present. The only contribution to the transverse part of the
cross section comes from the convection current. As a result, the
transverse response falls faster than the longitudinal response with
increasing momentum transfer, as will be shown explicitly below. This
is in contrast to the case of spin-1/2 quarks where the magnetization
current dominates. In turn, it causes the transverse response to
dominate at large momentum transfer, giving rise to the Callan-Gross
equation \cite{callangross} in the scaling region.

The inclusive, inelastic electron scattering cross section can be
re-expressed in terms of two structure functions, $W_1$ and $W_2$,
which depend only on $\nu$ and $Q^2$:
\begin{equation}
\frac{d \sigma} {d E_f d \Omega_f} = \sigma_{Mott}
\left ( W_2 (\nu, Q^2) + 2 \, W_1 (\nu, Q^2) \, \tan^2
\frac{\vartheta_e}{2} \right ) \,.
\label{defw}
\end{equation}
where
\begin{eqnarray}
W_1 (\nu, Q^2) & = &\frac{1}{2}R_T(\sqrt{Q^2+\nu^2},\nu)
\nonumber\\
W_2 (\nu, Q^2) & = & \frac{Q^4}{(Q^2+\nu^2)^2}
R_L(\sqrt{Q^2+\nu^2},\nu)+\frac{Q^2}{2(Q^2+\nu^2)}
R_T(\sqrt{Q^2+\nu^2},\nu) \,.
\label{eqw}
\end{eqnarray}

\section{The Coulomb Sum Rule}

For the moment, we will consider a wider class of models for hadrons
made up from confined quarks, namely the more general case of models
where all quarks carry an
electric
charge.  It is interesting to consider an
apparent contradiction between a model such as the one discussed here,
and the parton model.  In
our
model, since all states are bound
states, all of the transition form factors are coherent in that they
are the result of scattering from the total charge.  The parton model
however assumes that the cross sections are composed of incoherent
scattering from the individual constituents resulting in cross
sections proportional to the sum of squares of individual charges.
One method of examining the transition from coherent to incoherent
scattering is the Coulomb Sum Rule \cite{mcvoyvh}. Consider the
longitudinal response function
\begin{equation}
R_L(\vec q,\nu)=\sum_f < \psi_0|\rho^\dagger(\vec q)|\psi_f> <
\psi_f|\rho(\vec q)|\psi_0>\delta(\nu+E_0-E_f)
\end{equation}
where the sum represents a generalized sum over all final states,
bound or free, and $\rho(\vec q)$ is the Fourier transform of the
charge operator. Now define the logitudinal sum as
\begin{equation}
S(\vec q)=\int_{-\infty}^{\infty}d\nu R_L(\vec
q,\nu)\label{CoulombSum}
\end{equation}
Using the above definition of the longitudinal response and the
completeness of the final states, this becomes
\begin{eqnarray}
S(\vec q)&=&\int_{-\infty}^{\infty}d\nu \sum_f <
\psi_0|\rho^\dagger(\vec q)|\psi_f> < \psi_f|\rho(\vec
q)|\psi_0>\delta(\nu+E_0-E_f) \nonumber\\
&=&\sum_f < \psi_0|\rho^\dagger(\vec q)|\psi_f> < \psi_f|\rho(\vec
q)|\psi_0>\nonumber\\
&=&< \psi_0|\rho^\dagger(\vec q)\rho(\vec q)|\psi_0>\,.
\end{eqnarray}
This is a general result. To see how this relates to the problem
of the transition between coherent and incoherent scattering,
consider the simple case of a nonrelativistic system of two
constituents with charges $e_1$ and $e_2$. In this case
\begin{eqnarray}
S(\vec q)&=&\int d^3r\psi_0^\dagger(\vec r)\left( e_1 e^{-i\vec
q\cdot\vec r_1}+e_2 e^{-i\vec q\cdot\vec r_2}\right)\left( e_1
e^{i\vec q\cdot\vec r_1}+e_2 e^{i\vec q\cdot\vec r_2}\right)
\psi_0(\vec r)\nonumber\\
&=&\int d^3r\psi_0^\dagger(\vec r)\left( e_1^2 +e_2^2+e_1e_2
e^{-i\vec q\cdot(\vec r_1-\vec r_2)}+e_1e_2 e^{i\vec q\cdot(\vec
r_1-\vec r_2)}\right)
\psi_0(\vec r)\nonumber\\
&=&e_1^2+e_2^2+2e_1e_2{\cal F}(\vec q) \,,
\label{twoch}
\end{eqnarray}
where $\vec r=\vec r_1-\vec r_2$ is the relative coordinate of the
two particles and
\begin{equation}
{\cal F}(\vec q)=\Re \int d^3r\psi_0^\dagger(\vec r) \, e^{i\vec
q\cdot \vec r} \, \psi_0(\vec r)
\end{equation}
is the real part of the Fourier transform of the ground-state probability
density.

In order to understand the physical significance of the quantity
determining the rate of fall-off of the mixed term containing the
product of $e_1$ and $e_2$, it is necessary to write down the most
general form of the charge form factor for two quarks with charges
$e_1, e_2$ and masses $m_1, m_2$ ($M = m_1 + m_2$). Here, we have
dropped the $\delta$-function obtained from integrating over the
c.m. motion in the second step:
\begin{eqnarray}
F_{0,N_f} (\vec q ) &=& \int d^3 \vec r \,
 \Psi_{N_f}^* (\vec r) \, \left ( e_1 e^{i\vec q\cdot\vec r_1}+e_2
e^{i\vec q\cdot\vec r_2} \right) \Psi_0 (\vec r) \nonumber\\
& \rightarrow& \int d^3 \vec r \,
 \Psi_{N_f}^* (\vec r) \, \left ( e_1 e^{i\vec q\cdot\vec r \frac{m_2}{M}}
+e_2 e^{-i\vec q\cdot\vec r \frac{m_1}{M}} \right) \Psi_0 (\vec r)
\end{eqnarray}
From this expression one sees that in the most general case,
${\cal F}(\vec q)$ cannot be interpreted in terms of the
ground-state charge form factor. However, in the special case of
$m_1 = m_2$,
\begin{equation}
F_{0,0}(\vec q)=(e_1+e_2)\int d^3 \vec r \, e^{i\vec
q\cdot\frac{\vec r}{2}}|\Psi_0 (\vec r)|^2=(e_1+e_2)f_{0,0}(\vec
q)\,.
\end{equation}
In this case  \cite{frank}
\begin{equation}
{\cal F}(\vec q)=f_{0,0}(2\vec q) \,.
\end{equation}
A special case of our general result (\ref{twoch}) was discussed in
\cite{closeisgur}, where the case of two scalar, equal mass quarks in
a non-relativistic harmonic oscillator potential was considered.

In the model we present in this paper, all of the charge
is carried by one of the constituents, that is $e_1=1$ and
$e_2=0$. Since there is only a single charge, there is no
difference between coherent and incoherent scattering, so we expect
that
\begin{equation}
S(\vec q)=1\,.
\end{equation}
This then provides a useful test of the model.

Using (\ref{RL}),
\begin{eqnarray}
S(\vec q)&=&\int_{-\infty}^{\infty}d\nu \sum_{N=0}^\infty
\frac{1}{4 \, E_0 \, E_N}
 |F_{0N} (\vec q)|^2\nonumber\\
 &&\times\left[ (E_0+ E_N)^2 \delta(\nu + E_0 - E_N)-
 (E_0- E_N)^2 \delta(\nu + E_0 + E_N)\right]\nonumber\\
 &=&\sum_{N=0}^\infty
\frac{(E_0+ E_N)^2-(E_0- E_N)^2}{4 \, E_0 \, E_N}
 |F_{0N} (\vec q)|^2\nonumber\\
 &=&\sum_{N=0}^\infty |F_{0N} (\vec q)|^2
\end{eqnarray}
Using (\ref{eqff}) it is straightforward to demonstrate that $S(\vec
q)=1$ in this case so the Coulomb Sum Rule is satisfied.  Indeed, this
will be true regardless of the form of the confining potential, as
long as one considers a complete set of solutions. Note that for this
model it is necessary that the integral in (\ref{CoulombSum}) be over
both positive and negative energy transfers for the sum rule to be
satisfied.

\begin{figure}
\epsfxsize=5in \centerline{\epsffile{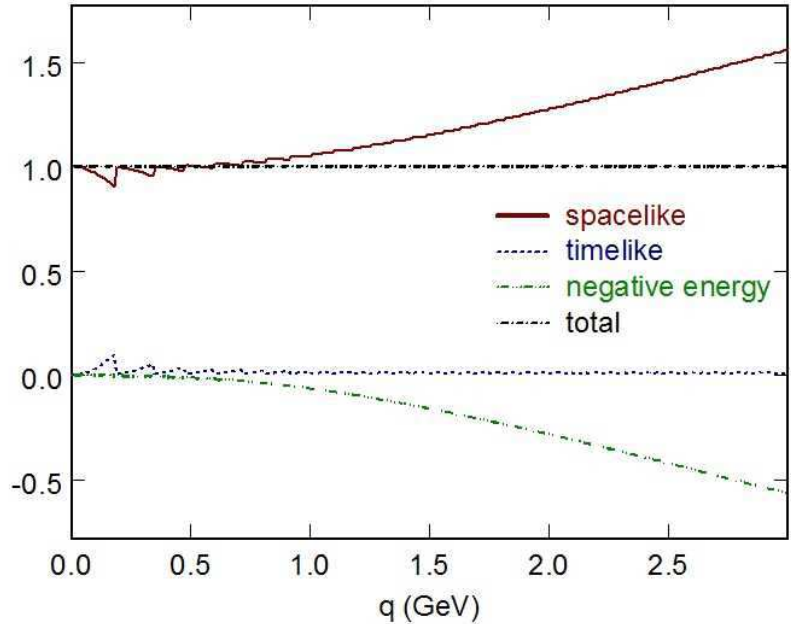}}
\caption{Positive energy spacelike, timelike and negative energy
contributions to the Coulomb Sum rule from our model. }\label{CSRplot}
\end{figure}

In electron scattering, only the spacelike region is accessible and
the negative energy states are an artifact of the use of the
Klein-Gordon equation as a wave equation. It is useful, therefore, to
examine the contributions to the Coulomb sum from spacelike, timelike
and negative energy states. When referring to the spacelike and
timelike contributions, only positive energy states are included.  The
different contributions are shown in Fig. \ref{CSRplot}. At $|\vec
q|=0$, only the elastic form factor can contribute and therefore must
saturate the sum rule. As the momentum transfer increases the elastic
form factor falls off, resulting in a decrease in the spacelike
contribution. It decreases until the momentum transfer increases to a
point where the first excited state enters the spacelike region. The
spacelike contribution then saturates the sum rule again. This process
continues as new form factors become accessible in the spacelike
region. The result is a saw-toothed behavior of the spacelike
contribution. Because the density of states for this oscillator model
increases with increasing energy, the magnitude of the ``teeth''
becomes smaller with increasing momentum transfer until the spacelike
contribution is essentially smooth.  Note that the spacelike
contribution over-saturates the sum rule around momentum transfers
of 0.5 GeV and above.
Since the jaggedness in the spacelike region is associated with
the migration of contributions from the timelike region, it is not
surprising to see that the complement of this behavior does indeed
show up in the timelike contribution. As the momentum transfer
increases, the size of the timelike contribution becomes small.
The negative energy contribution is smooth and compensates for the
over-saturation of the spacelike contribution. This is clearly an
artifact of using a one-body wave equation with negative energy
contributions.

\begin{figure}
\epsfxsize=5in \centerline{\epsffile{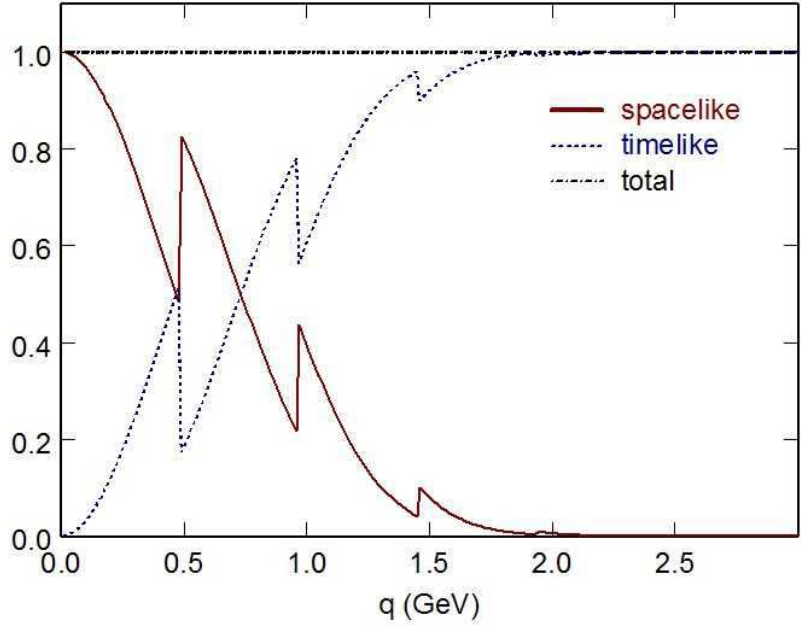}}
\caption{Spacelike and timelike contributions to the Coulomb Sum
Rule for a nonrelativistic oscillator.}\label{CSRplotNR}
\end{figure}

It should be pointed out here that the small size of the timelike
contributions is an essential consequence of using a relativistic wave
equation.  This can be seen by considering a similar situation where
the nonrelativistic oscillator model is used to describe the system.
The spacelike and timelike contributions of such a model are shown in
Fig. \ref{CSRplotNR}. The Schr\"odinger equation of course has only
positive energy solutions.  Here the saw-toothed behavior in the
spacelike contributions is similar to the relativistic case with the
important difference that due to the linear character of the
nonrelativistic spectrum (see Fig.~\ref{figenercomp}) the
contributions from states entering the spacelike region is not rapid
enough to compensate for the fall-off in the form factors.  Therefore,
with increasing momentum transfer, the size of the spacelike
contribution approaches zero with all of the strength appearing in the
timelike region.

Duality in the sum of the form factors in the spacelike region is
related to the Coulomb sum rule. While the concept of duality in the
form factors is not clearly related to an observable (in contrast to
 duality in
the structure functions), it still has received attention in the
literature \cite{closeisgur}.  We describe it within our model in
appendix \ref{ffapp}.

This model can be easily extended to the case where both quarks
are charged by examining the behavior of the two-body Gross
equation in the limit where the mass of one of the particles
becomes infinite \cite{waw}. The
contribution of the infinite mass particle to the structure
function is simple and straight forward.  Due to its infinite
mass, this particle remains stationary for any finite momentum
transfer, and it is point-like. This particle therefore contributes
only to elastic scattering and has a constant form factor. The
structure function can then be written as
\begin{eqnarray}
R_L(\vec q,\nu)&=& |e_1 F_{00} (\vec q)+e_2|^2\nonumber\\
&&+e_1^2 \sum_{N=1}^\infty \frac{1}{4 \, E_0 \, E_N}
 |F_{0N} (\vec q)|^2\left[ (E_0+ E_N)^2 \delta(\nu + E_0- E_N)
 \right.\nonumber\\
 &&\qquad\qquad -\left.
 (E_0- E_N)^2 \delta(\nu + E_0 + E_N)\right]\,\label{RL2charge}
\end{eqnarray}
The Coulomb Sum can be easily calculated to be
\begin{eqnarray}
S(\vec q)&=&e_1^2\sum_{N=0}^\infty |F_{0N}(\vec q)|^2+e_2^2+2 e_1
e_2 \Re F_{00}(\vec q)\nonumber\\
&=& e_1^2+e_2^2+2e_1e_2\Re F_{00}(\vec q)
\end{eqnarray}
So in this case
\begin{equation}
{\cal F}(\vec q)=\Re F_{00}(\vec q)=F_{00}(\vec q)\,.
\end{equation}
After examining how the apparent contradiction between coherent and
 incoherent scattering is resolved in a more general framework, we now
proceed to investigate duality in our model. The first condition for duality
is that one obtains scaling in the structure function calculated solely with
resonances, and that the scaling curve thus obtained agrees with the
scaling curve obtained in the parton model.

\section{The Parton Model}
\label{secpm}

The usual assumption of the parton model is that at large momentum
transfers the final state quarks can be treated as though they were
free. Examination of the structure functions for large $Q^2$ and
fixed Bjorken $x$ leads to identification of the scaling
functions.

For our simple model, the response functions for excitation of a
bound (off-mass-shell) quark to a plane-wave final state can be
calculated analytically as
\begin{eqnarray}
R_L(\vec q,\nu) &=& \frac{(2E_0+\nu)^2}{4\pi^\frac{1}{2}\beta E_0
|\vec q|}\left[ \exp{\left(-\frac{y^2}{\beta^2}\right)}-
\exp{\left(-\frac{(y+2|\vec
q|)^2}{\beta^2}\right)}\right]\nonumber\\
&&\times \left[ \theta(\nu+E_0-m)-\theta(-\nu-E_0-m)\right]
\end{eqnarray}
and
\begin{eqnarray}
R_T(\vec q,\nu) &=& \frac{\beta}{2\pi^\frac{1}{2} E_0 |\vec
q|^2}\left[ \left( 2(y+|\vec q|)-\frac{\beta^2}{|\vec q|}\right)
\exp{\left(-\frac{y^2}{\beta^2}\right)}\right.\nonumber\\
&&-\left.  \left( 2(y+|\vec q|)+\frac{\beta^2}{|\vec q|}\right)
\exp{\left(-\frac{(y+2|\vec
q|)^2}{\beta^2}\right)}\right]\nonumber\\
&&\times \left[ \theta(\nu+E_0-m)-\theta(-\nu-E_0-m)\right]
\end{eqnarray}
where $y=\sqrt{(\nu+E_0)^2-m^2}-|\vec q|$.

For our model it is not possible to define the scaling variable in
terms of the target mass since it is infinite. For this reason we
define a new Bjorken variable
\begin{equation}
u_{Bj} = \frac{M}{m} x_{Bj}=\frac{Q^2}{2m\nu}
\end{equation}
which covers the interval $-\infty<u_{Bj}<\infty$. Using
$\nu=\frac{Q^2}{2mu_{Bj}}$,
and taking the limit $Q^2\rightarrow\infty$, the structure
functions become
\begin{equation}
R_L(Q^2,u_{Bj})  \rightarrow \frac{Q^2}{8\pi^\frac{1}{2}\beta
mE_0u_{Bj}} \exp{\left(-
      \frac{(E_0-mu_{Bj})^2}{\beta^2}\right)}
      \left[\theta(\frac{Q^2}{2mu_{Bj}})-\theta(-\frac{Q^2}{2mu_{Bj}})\right]
 \,,\label{RLPWscale}
\end{equation}
and
\begin{equation}
R_T(Q^2,u_{Bj})  \rightarrow \frac{2\beta
mu_{Bj}}{\pi^\frac{1}{2}E_0Q^2} \exp{\left(-
      \frac{(E_0-mu_{Bj})^2}{\beta^2}\right)}
      \left[\theta(\frac{Q^2}{2mu_{Bj}})-\theta(-\frac{Q^2}{2mu_{Bj}})\right]
 \,.\label{RTPWscale}
\end{equation}
Since in this limit
\begin{equation}
|\nu|\frac{Q^4}{\vec q ^4}\rightarrow \frac{8m^3|u_{Bj}|^3}{Q^2}
\mbox{ and }
|\nu|\frac{Q^2}{2\vec q ^2}\rightarrow m|u_{Bj}|
\end{equation}
the structure functions have the limits
\begin{equation}
W_1(u_{Bj},Q^2)\rightarrow F_1(u_{Bj})=0
\end{equation}
and
\begin{equation}
|\nu| W_2(\nu, Q^2)  \rightarrow F_2(u_{Bj})=\frac{m^2
u_{Bj}^2}{\pi^\frac{1}{2}\beta E_0} \exp{\left(-
      \frac{(E_0-mu_{Bj})^2}{\beta^2}\right)}
 \,.
\label{sfanalyt}
\end{equation}
Note that the choice of $|\nu|$ in defining  $W_2$ is necessary to
provide a properly normalized scaling function as will be seen
below.

Although we have used the Bjorken scaling variable to obtain these
results, this will be true for all such variables since all
acceptable scaling variables must reduce to the Bjorken scaling
variable as $Q^2\rightarrow\infty$. Therefore, a more general
expression for $F_2$ for any scaling variable and any initial
state can be written as
\begin{equation}
F_2(u)=\frac{m^2u^2}{4\pi^2E_0}\int_{|E_0-mu|}^\infty
dp \, p N(p)\label{F2momdist}
\end{equation}
where $N(p)$ is the ground state momentum distribution normalized
such that
\begin{equation}
\frac{1}{2\pi^2}\int_0^\infty dp p^2 N(p)=1 \, .
\end{equation}
After obtaining the scaling curve in the parton model, i.e. the scaling
curve for a quark initially bound and then free, we proceed to find an
expression for the scaling curve in our model, where the quark makes the
transition from the ground stated to an excited bound state.

\section{Continuum Limit}
\label{seccl}

An interesting feature of our relativistic oscillator model is
that the scaling behavior of the model can be determined
analytically by making a continuum approximation. The
justification for this is that at increasing momentum transfer the
contributions to the response functions are dominated by higher
energy states. Since the density of states increases with
increasing energy, it is reasonable that a continuum approximation
should provide a good description of the averaged response for
large momentum transfers.

Using (\ref{RL}) and ({\ref{eqff}) we can write
\begin{eqnarray}
R_L(\vec q,\nu) &=& \sum_{N = 0}^{\infty} \,\Delta N \frac{1}{4
E_0 E_N} \, \, \frac{1}{N!}  \left (\frac{\vec q \, ^2} {2
\beta^2}
\right )^N \exp{(-\frac{\vec q \, ^2}{2 \beta^2})} \nonumber\\
&&\times \left[(E_N+E_0)^2\delta(E_N - E_0 -
\nu)-(E_0-E_N)^2\delta(E_N +E_0 + \nu)\right]
\end{eqnarray}
where $\Delta N=1$. It is convenient to write
\begin{equation}
E_N=\sqrt{p_N^2+E_0^2}
\end{equation}
where
\begin{equation}
p_N^2=2\beta^2N\,.
\end{equation}
>From this it can be determined that for a variation in $N$,
\begin{equation}
\Delta N=\frac{p_N\Delta p_N}{\beta^2}\,,
\end{equation}
and
\begin{equation}
\Delta E_N=\frac{p_N\Delta p_N}{E_N}\,.
\end{equation}
The longitudinal response function can then be rewritten as
\begin{eqnarray}
R_L(\vec q,\nu) &=& \frac{1}{4\beta^2E_0}\sum_{N = 0}^{\infty} \,
\frac{\Delta E_n}{\Gamma\left(
1+\frac{E_N^2-E_0^2}{2\beta^2}\right)} \left (\frac{\vec q \, ^2}
{2 \beta^2} \right )^\frac{E_N^2-E_0^2}{2\beta^2}
\exp{(-\frac{\vec q \, ^2}{2 \beta^2})}\nonumber\\
&&\times \left[ (E_N+E_0)^2\delta(E_N - E_0 -
\nu)-(E_0-E_N)^2\delta(E_N + E_0 + \nu)\right] \,.
\end{eqnarray}
This sum can now be approximated by the integral
\begin{eqnarray}
R_L(\vec q,\nu) &=& \frac{1}{4\beta^2E_0}\int_{E_0}^{\infty} \,
\frac{dE }{\Gamma\left( 1+\frac{E^2-E_0^2}{2\beta^2}\right)} \left
(\frac{\vec q \, ^2} {2 \beta^2} \right
)^\frac{E^2-E_0^2}{2\beta^2} \exp{(-\frac{\vec q
\, ^2}{2 \beta^2})} \nonumber\\
&&\times\left[(E+E_0)^2 \delta(E - E_0 - \nu)-(E-E_0)^2 \delta(E +
E_0 + \nu)\right]\,
\end{eqnarray}
which can be trivially evaluated to give
\begin{equation}
R_L(\vec q,\nu) = \frac{(\nu+2
E_0)^2\left[\theta(\nu)-\theta(-2E_0-\nu)\right]}{4\beta^2E_0
\Gamma\left( 1+\frac{(\nu+E_0)^2-E_0^2}{2\beta^2}\right)} \left
(\frac{\vec q \, ^2} {2 \beta^2} \right
)^\frac{(\nu+E_0)^2-E_0^2}{2\beta^2} \exp{(-\frac{\vec q \, ^2}{2
\beta^2})} \,. \label{RLconta}
\end{equation}
Similarly, the transverse response becomes
\begin{equation}
R_T(\vec q,\nu) = \frac{(\nu^2+2E_0\nu
)\left[\theta(\nu)-\theta(-2E_0-\nu)\right]}{E_0 \vec
q^2\Gamma\left( 1+\frac{(\nu+E_0)^2-E_0^2}{2\beta^2}\right)} \left
(\frac{\vec q \, ^2} {2 \beta^2} \right
)^\frac{(\nu+E_0)^2-E_0^2}{2\beta^2} \exp{(-\frac{\vec q \, ^2}{2
\beta^2})} \,. \label{RTconta}
\end{equation}

In the scaling limit the argument of the $\Gamma$ function becomes
large, so Stirling's formula can be used to write the longitudinal
response function as
\begin{eqnarray}
R_L(\sqrt{ Q^2+\nu^2},\nu) &=& \frac{(\nu+2
E_0)^2}{4\pi^\frac{1}{2}\beta E_0} \frac{\exp{\left(
      \frac{\nu^2+2E_0\nu}{2\beta^2}\ln\left(\frac{Q^2+\nu^2}
{\nu^2+2E_0\nu}\right)-\frac{Q^2+\nu^2}{2\beta^2}+\frac{\nu^2+2E_0\nu}
{2\beta^2}\right)}}{\sqrt{\nu^2+2E_0\nu}}\nonumber\\
&&\times\left[\theta(\nu)-\theta(-2E_0-\nu)\right]\,.\label{wcontb}
\end{eqnarray}
Using
\begin{equation}
\nu=\frac{Q^2}{2mu_{Bj}}\,,
\end{equation}
and taking the limit $Q^2\rightarrow\infty$, the  structure
functions become
\begin{equation}
R_L(Q^2,u_{Bj})  \rightarrow \frac{Q^2}{8\pi^\frac{1}{2}\beta
mE_0u_{Bj}} \exp{\left(-
      \frac{(E_0-mu_{Bj})^2}{\beta^2}\right)}
      \left[\theta(\frac{Q^2}{2mu_{Bj}})-\theta(-\frac{Q^2}{2mu_{Bj}})\right]
 \,,\label{RLContScale}
\end{equation}
Similarly,
\begin{equation}
R_T( Q^2,u_{Bj})  \rightarrow \frac{2\beta
mu_{Bj}}{\pi^\frac{1}{2}E_0Q^2} \exp{\left(-
      \frac{(E_0-mu_{Bj})^2}{\beta^2}\right)}
      \left[\theta(\frac{Q^2}{2mu_{Bj}})-\theta(-\frac{Q^2}{2mu_{Bj}})\right]
 \,.\label{RTContScale}
\end{equation}
Since (\ref{RLContScale}) and (\ref{RTContScale}) are identical to
(\ref{RLPWscale}) and (\ref{RTPWscale}) the model scales to the parton
model result. So, even though our model describes a bound quark being
excited to resonance states, we do obtain a scaling curve in the
Bjorken limit. This, as well as the results presented in
\cite{greenberg,gurvitzrinat,psl,pp}, show that scaling does not
necessarily imply scattering off free constituents, a belief which is
encountered widely.

 While others \cite{greenberg,gurvitzrinat,psl} have studied the
transition from ground state to excited bound states and found scaling
in similar models.
However,
 duality is fulfilled only when i) the transition
from ground state to excited bound states scales, ii) the transition
from ground state to a plane-wave final state scales, {\it and} iii) both
scaling curves coincide. In this section and the preceding section, we
have shown that duality holds explicitly in our model. The numerical approach
towards the scaling curve is shown in Fig.~\ref{figscal1}.

Note that as we explicitly made use of Stirling's formula in the
derivation of the scaling function in the continuum limit, it is clear
that for the lower lying resonances, which correspond to lower $N$
values, we will never quite see scaling in the subasymptotic regime. This
is of no practical relevance, as these resonances are pushed out to
very high values of $u$ for larger $Q^2$, and the structure functions
practically vanish in this region. This is completely analogous to the
fact that for electron scattering from a proton, one always picks up
the elastic scattering at $x_{Bj} = 1$, independent of $Q^2$ - even
though the elastic form factor will have fallen off to negligible
values at high enough $Q^2$.

\section{Approach to scaling in the structure functions}

After establishing analytically that one of the necessary conditions
for duality is fulfilled, namely scaling to the scaling curve
obtained from a free quark in the final state, we proceed to investigate
the approach to scaling numerically.

We would like to remind the reader that our results should not be
compared to the available nucleon data - our model calculations
describe a process that might resemble electron scattering from a $B$
meson, which has never been measured. In general, when we consider
scattering from a meson target, scaling will set in later than for a
baryon target: momentum sharing for higher momenta between fewer
constituents is easier, which leads to a slower fall-off of the
individual form factors, and to a later onset of scaling. In our
situation, where the system is not allowed to decay, we have a
somewhat extreme case.

   To see duality clearly both experimentally and theoretically, one
needs to go beyond the Bjorken scaling variable $x_{Bj}$ and the
scaling function ${\cal{S}}_{2Bj} = \nu W_2$ that goes with it.
This is because in deriving Bjorken's variable and scaling
function, one not only assumes $Q^2$ to be larger than any mass
scale in the problem, but also that high $Q^2$ (pQCD) dynamics
controls the interactions. However, duality has its onset in the
region of low to moderate $Q^2$, and there masses and violations
of asymptotic freedom do play a role. Bloom and Gilman used a new,
{\it ad hoc} scaling variable $\omega'$ \cite{bgduality} in an
attempt to deal with this fact. In most contemporary data
analyses, the Nachtmann variable \cite{greenbergb,nachtmann} is
used together with ${\cal{S}}_{2Bj}$. Nachtmann's variable
contains the target mass as a scale, but neglects quark masses.
For our model, the constituent quark mass (assumed to arise as a
result of spontaneous chiral symmetry breaking) is vital at low
energy, and a scaling variable that
does not make any assumptions about the size of the quark and
target masses compared to $Q^2$
is desirable. Such a variable was derived more than twenty
years ago by Barbieri {\it et al.}  \cite{barbieri} to take into
account the masses of heavy quarks; we use it here given that
after spontaneous chiral symmetry breaking the nearly massless
light quarks have become massive constituent quarks, calling it
$x_{cq}$:
\begin{equation}
x_{cq} = \left\{ \begin{array}{ll}
\frac{1}{2 M} \left(
\sqrt{\nu^2 + Q^2} - \nu \right) \left( 1 + \sqrt{1 + \frac{4
m^2}{Q^2}} \right)& \mbox{for $\nu >0$}\\
- \frac{1}{2 M}\left ( \sqrt{\nu^2 + Q^2} + \nu \right ) \left( 1
+ \sqrt{1 + \frac{4 m^2}{Q^2}} \right)& \mbox{for $\nu <0$}
\end{array}
\right.
\, . \label{defxdis}
\end{equation}
where the definition for negative energy is chosen such that it
satisfies the kinematic constraints in this region and reproduces
the behavior of $x_{Bj}$ for large $Q^2$. The scaling function
associated with this variable is given by:
\begin{equation}
\label{S}
  {\cal{S}}_{2cq} \equiv |\vec q|\  W_2 = \sqrt{\nu^2 + Q^2}\  W_2\,.
\end{equation}
This scaling function and variable were derived for scalar quarks
which are free, but have a momentum distribution. The derivation
of a new scaling variable and function for bound quarks will be
published elsewhere. Numerically, this scaling variable does not
differ very much from the one in Eq. (\ref{defxdis}). Of course
all versions of the scaling variable must converge to $x_{Bj}$ and
all versions of the scaling function must converge towards
${\cal{S}}_{Bj}$ for large enough $Q^2$. One can also easily
verify that in the limit $m \to 0$ one obtains from
(\ref{defxdis}) the Nachtmann scaling variable. In the following,
we use the variable $x_{cq}$ and the scaling function
${\cal{S}}_{2cq}$.

\begin{figure}[!h,t]    
\begin{center}
\epsfxsize=7in \centerline{\epsffile{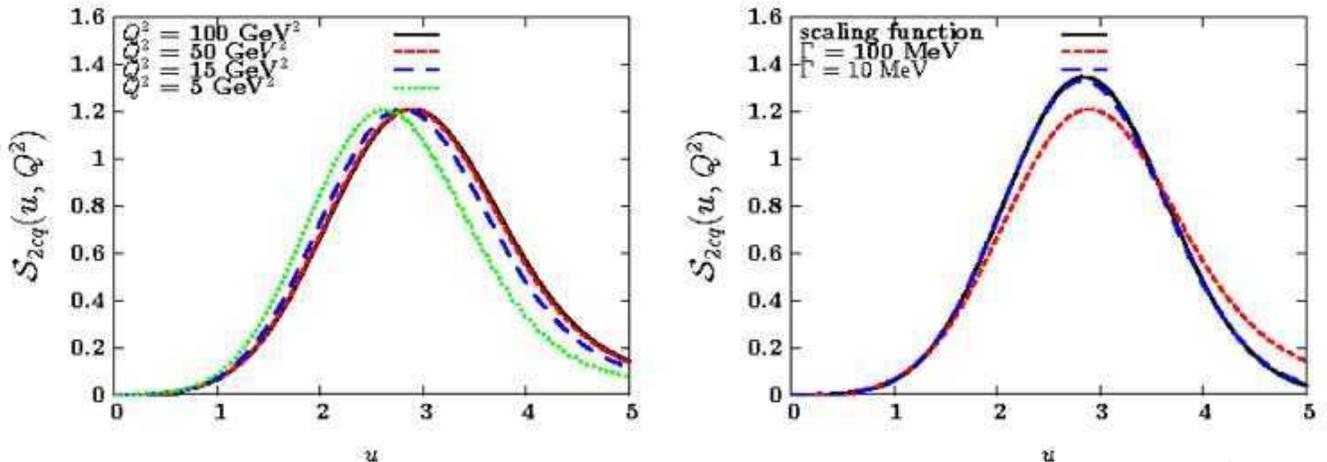}}
\end{center}
\vspace{-1.5cm}
\caption{The high energy scaling behavior of ${\cal{S}}_{2cq}$ as a
function of $u$ for various values of $Q^2$. In the left panel, we have used
$\Gamma = 100$~MeV to give the impression of real resonances even
though this large value distorts the scaling curve somewhat. For any
width equal to or smaller than this, the distortion is rather
innocuous, and for $\Gamma \rightarrow 0$, the structure function
approaches the scaling function (solid line)
 in Eq.~(\ref{sfanalyt}), as shown in
the right hand panel. The structure functions in the left panel are
shown for $Q^2$ = 100 GeV$^2$.}
\label{figscal1}
\end{figure}
  We are now ready to look at scaling and duality in our model. Since the
target has mass
$M
\rightarrow
\infty$, it is convenient to rescale the scaling variable $x_{cq}$   by a
factor
$M/m$:
\begin{eqnarray}
u & \equiv & {M \over m}\ x_{cq}\ ~~.
\label{defu}
\end{eqnarray}
The variable $u$ takes values from 0 to a maximal, $Q^2$
dependent value, which can go to infinity.
The high energy scaling behavior of the appropriately rescaled structure
function
${\cal {S}}_{cq}$ is illustrated in Fig.~\ref{figscal1}.
The structure function has been evaluated using the
phenomenologically reasonable parameters $m = 0.33$~GeV and $\beta
= 0.4 ~{\rm GeV}$. To display it in a visually meaningful manner,
the energy-dependent $\delta$-function has been smoothed out by
introducing an  unphysical Breit-Wigner shape with an arbitrary
but small width, $\Gamma$, chosen for purposes of illustration:
\begin{equation}
\delta(E_N - E_0 - \nu) \rightarrow \frac{\Gamma}{2 \pi} \, \,
\frac{f}{(E_N - E_0 - \nu)^2 + (\Gamma/2)^2}\, ,
\end{equation}
where the factor
$
f = {\pi}/[{\frac{\pi}{2}
                + \arctan {2 (E_N - E_0) \over \Gamma}] }
$
ensures that the integral over the $\delta$-function is identical
to that over the Breit-Wigner shape.
As for the all scalar case discussed in \cite{dp1}, the smearing out
of the $\delta$-function in energy with the Breit-Wigner shape leads to
a slight widening of the curve and flattening of the peak height. However,
when choosing a smaller value for the Breit-Wigner width, these effects
disappear, as seen in the right panel of Fig.~\ref{figscal1}.

Not unexpectedly, the scaling curve we find here, when using photons
and electrons with their appropriate spins, differs from the one in
the all scalar case both in its final shape and the approach to
scaling.  Now, we have two terms, the longitudinal and transverse
response function, contributing to the structure function $W_2$ and
therefore to the scaling curve. More importantly, the terms themselves
are different and more complicated in the case considered here.  The
longitudinal part of the structure function contains an additional
factor $(E_0 + E_N)^2$, which was not present in the all scalar case.
As shown in section \ref{seccl}, the transverse response vanishes like
$1/Q^2$ in the limit $Q^2 \rightarrow \infty$, while the longitudinal
response grows like $Q^2$.  This leads to a vanishing of $W_1$, and to
a $Q^2$ independent value for $|\vec q| W_2$. Even though, from
eqs.(\ref{eqw},\ref{RLContScale},\ref{RTContScale}) it is clear that
at lower $Q^2$, the transverse contribution to $W_2$ will not vanish
immediately, therefore making the approach to scaling slower.  The
effect is rather significant, though, as the contributions of the
convection current are very small. They delay scaling slightly for the
low $u$ part of the curve.  The effect would be more important for
contributions of similar size within a certain kinematic range.
In our case, the main effect of the transverse contribution is to slightly
broaden the curve. For smaller $Q^2$, this effect is more pronounced for
low values of the scaling variable $u$, as the higher $u$ correspond
to lower lying resonances, which have only tiny contributions from the
transverse part.

 As already mentioned above, for a proton target, the dominant
contribution to the transverse response and overall is the
magnetization current, which does not contribute for our scalar
``quarks''.
Note that both the transverse and longitudinal contribution to
$W_2$ are positive definite.
If a dominant contribution in the transverse response is
present, it should lead to a different scaling behavior
 in the structure function $W_1$ than in $\nu W_2$.
For  $\nu W_2$, the longitudinal term with different
$Q^2$ behavior will most likely slow the approach to
scaling down, as it is going to be of
comparable size to the magnetization current contribution at low
$Q^2$.
This is a completely general observation, and one would expect
to see faster scaling in $F_1$ once the data are available.  The same
conclusion was reached on a different basis in \cite{closeisgur}.

The shape of the scaling curve is also different than for the all
scalar case.  The peak is higher, the curve extends to larger values
of the scaling variable, and for $ u \rightarrow 0$, the scaling
function actually vanishes now, as expected from a valence quark
distribution, even though we do not find a behavior $\propto \sqrt{u}$
as seen for proton targets. However, we cannot expect to reproduce the
correct distribution function for quarks in our simple model with
scalar ``quarks''.

From the explicit expression for the scaling curve, it is clear that
it depends both on the binding strength $\beta$ of the harmonic
oscillator, and on the quark mass. It peaks at $ u_{peak} = (E_0 +
\sqrt{E_0^2 + \beta^2})/2 m$, slightly above the value
$u_{peak}^{scalar} = \frac{E_0}{m}$ which we found for the all scalar
case. Naively, for a target of mass $M$ made up of non-interacting
quarks of mass $m_q$, one expects a spike at $x_{Bj} =
\frac{m_q}{M}$. In our case, the role of the mass of the quark $m_q$
is played by the ground state energy $E_0$, which appears everywhere
(e.g. in flux factors, normalization) where one would have the mass in
the free particle case. As our variable is rescaled with the factor
$\frac{M}{m}$, we expect $u_{peak} \approx \frac{E_0}{m}$. It is
interesting to note that this value receives a slight binding
correction due to the conserved current employed here.  We note that
for weaker binding, the peak gets narrower and its position slides
towards $u \approx 1$. As expected, in the limit of a free particle, $\beta
\rightarrow 0$, we do obtain a spike of infinite height at $u =1$ [in this
limit, the scaling function becomes $\delta(u)/(m E_0)$].

\section{Moments and further Sum Rules}

Now, we will discuss global duality, where the term global implies
that we consider an average/integral over many resonances, which
is compared with the corresponding integral over the scaling curve.

Local duality implies that we compare the contribution of one single
resonance or just a few resonances with the scaling result, i.e. with
the free quark result. This will be discussed in section
\ref{secloc}.  The concept of local duality is taken to its extreme
when one focuses not just on one single resonance, but on one point
only of the contribution of the single resonance, as it was done first
by Bloom and Gilman \cite{bgduality}, when they compared the peak value
of a single resonance with the value of the scaling curve.  This
version of duality has been investigated in Ref.~\cite{carl}. As this
ratio would depend strongly on the Breit-Wigner width we use to
smooth out the $\delta$-functions, it is not appropriate to consider
it in this paper.

Global duality was first quantified by Bloom and Gilman \cite{bgduality}
in the form of finite energy sum rules, where the integral over the
scaling curve was compared to the integral over the resonance
contribution. The integration range in both cases comprises the region
of the scaling variable $\omega'$ or $\nu$, respectively, which
corresponds to the resonance region, defined as having an invariant mass
$W < 2$ GeV:
\begin{equation}
\frac{2 M_N}{Q^2} \int_0^{\nu_m} d \nu \nu W_2 (\nu, Q^2)
=
\int_{1}^{1+W_m^2/Q^2} d \omega' \nu W_2 (\omega') \,,
\end{equation}
where $W_m \approx 2 $ GeV, and $\nu_m = \frac{W_m^2 - M_N^2 + Q^2}{2
M_N}$.  Here, $M_N$ denotes the mass of the nucleon target.  The
agreement between the left and right hand side of this equation is
better than $10~\%$; for the larger values of Q$^2$, starting around
Q$^2 \approx 2$ GeV$^2$,
the agreement is
quite impressive: $ 2 \%$ or better.

While it certainly would be desirable to calculate the same finite
energy sum rule in our model, there is a practical problem
and a philosophical
problem. Firstly, in our model, we deal with an infinitely heavy
system, so that in principle, the invariant mass $W$ of the final
state is always infinity. Even if we could define a reasonable
substitute for the invariant mass of the final state, picking an
integration limit is a problem in principle: for our model, the
scaling curve consists solely of resonance contributions, even though
they cannot be resolved and form a smooth curve. So, any distinction
between ``resonance region'' and ``continuum'' or ``multiparticle
final states'' is artificial. The conventional definition of
``resonance region'' as the region where $W < 2$ GeV means the region
where the resonances are prominent and dominant. However, it does not
mean that for $W > 2$ GeV, there are no resonances present, and it
also does not mean that for $W < 2$ GeV, i.e. in the resonance region,
there are no non-resonant contributions at all. Experimentally, there
is background for $W < 2$ GeV, and there are resonances for $W > 2$
GeV, e.g. several $N^*$ and $\Delta$ resonances. Also from a
theoretical point of view, it is obvious that resonances which decay
by creation of a quark-antiquark pair in the final state must be
accompanied by a corresponding non-resonant production mechanism,
where the pair creation takes place in the initial state, and the
photon interacts with the preformed meson.

Since the distinction between a ``resonance region'' and a ``continuum
region'' has its problems, we utilize the moments of the scaling
function ${\mathcal{S}}_{cq}$, where the integration range comprises the
whole interval of the scaling variable. The physical information
contained in the finite energy sum rule and the moments $M_n$ is the
same. The moments of a scaling function ${\mathcal{S}} (Q^2, x)$ with
a scaling variable $x$ are defined as
\begin{equation}
M_n (Q^2) = \int_0^{x_{max}} dx \, \, x^{n-2} \, {\mathcal{S}}_2
(Q^2, x)\, .
\end{equation}
Here, in contrast to \cite{dgp}, we do not include the unphysical
region $]x_{max},1]$ in the integration interval.  It is obvious that
higher moments, i.e. $n = 4,6,\dots$, tend to emphasize the resonance
region, as for fixed Q$^2$, the resonances are found at large $x$.
The values of the moments decrease with increasing $n$.  In our case,
we change to $u$-type scaling variables, see Eq.~\ref{defu}, so that
\begin{equation}
M_n^{u-based} (Q^2) = \int_0^{u_{\rm max}} du \, \, u^{n-2} \,
{\mathcal{S}}_2 (Q^2, u)\, ,
\end{equation}
where $u_{\rm max}$ corresponds to the maximum value of $u$ which is
kinematically accessible at a given $Q^2$.
By changing from $x$ to $u$ scaling variables, we change the upper
integration limit from a value equal to or lower than 1 to a value
considerably larger than 1 for Q$^2 > 1 - 2 $ GeV$^2$. This means
that our higher moments will emphasize the low-lying resonances
even more than the conventional, $x$-based moments. Also, the
higher moments will be larger than the moments with small $n$.

 Evaluating the moments of the structure function (\ref{S})
explicitly  one has
\begin{eqnarray}
M_n(Q^2) &=& \left ( \frac{r}{2 m} \right )^{n-1} \,
\sum_{N=0}^{\infty} \, \left( \sqrt{\nu_N^2 + Q^2} - \nu_N
\right)^{n-1} \, \, \frac{E_0}{ E_N} \, \left|F_{0N}
\left(\sqrt{\nu_N^2 + Q^2} \right) \right|^2 ~~\nonumber\\
& &\qquad\qquad\times \left[ \frac{Q^4}{q_N^4} (E_0+ E_N)^2 + 4 N
\alpha \frac{Q^2}{q_N^4} \right ] \,,\label{moments}
\end{eqnarray}
with $r \equiv 1 + \sqrt{1 + 4 m^2/Q^2}$, and $\nu_N = E_N - E_0$ and
$q_N = \sqrt{Q^2 + \nu_N^2}$.

The elastic contribution
is given by
\begin{equation}
M_n ^{elastic} (Q^2) = \left ( \frac{r}{2 m} \right )^{n-1} \,
|Q|^{n-1} \, \exp (-\frac{Q^2}{2 \beta^2})
\end{equation}
Note that for vanishing four-momentum transfer $Q^2$, all moments
take the value $1$, independent of $n$.

\begin{figure}[!h,t]
\begin{center}
\epsfxsize=5in \centerline{\epsffile{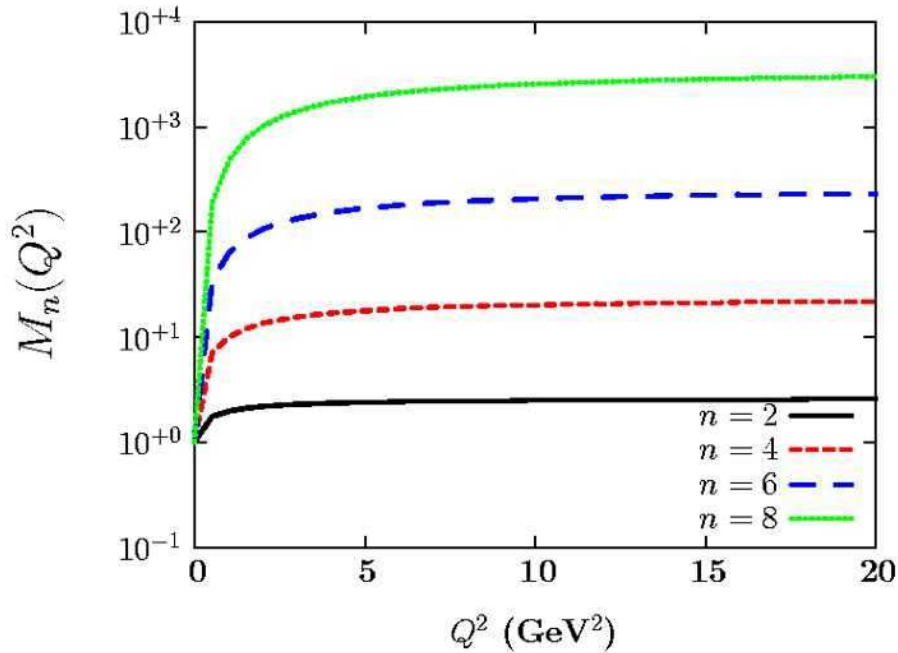}}
\end{center}
\vspace{-0.5cm}
\caption{The lowest moments $M_n$ as a function of $Q^2$.}
\label{figmom}
\end{figure}

In Fig.~\ref{figmom}, we show the moments for $n = 2,4,6,8$, which
were obtained by integrating over the positive energy states only.
One can see clearly that all moments flatten out, even though they
did not quite reach their asymptotic value at the highest $Q^2$
value shown.  The lowest moment $M_n$, is within 9~\% of its
asymptotic value at $Q^2 = 5$ GeV$^2$, and within ~ \% of its
asymptotic value at $Q^2 = 20$ GeV$^2$. As expected, the higher
moments, which by construction get more contributions from the
lower lying resonances, need higher $Q^2$ values in order to reach
their asymptotic values. For $M_6$, we find that it has reached 64
\% of its asymptotic value at $Q^2 = 5$ GeV$^2$, and 88~\% of its
asymptotic value at $Q^2 = 20$ GeV$^2$. From these numbers, we can
see that even though scaling does not set in for $Q^2<50$~GeV$^2$,
the asymptotic values at least of the lower moments are reached
much earlier. This reflects the fact that ${\cal{S}}_{cq}(u,Q^2)$
approaches the scaling curve by shifting towards higher $u$, not
by approaching it from below or above.

Since the continuum approximation provides a relatively simple
analytic expression for the structure functions, it is possible to
use this to study certain properties of the moments. First,
however, it is necessary to determine the validity of this
approximation for the calculation of moments. Figure
\ref{contmoments} shows calculations of the first three moments
$M_0$, $M_1$ and $M_2$.
\begin{figure}
\vspace{-0.5cm}
\epsfxsize=3.75in \centerline{\epsffile{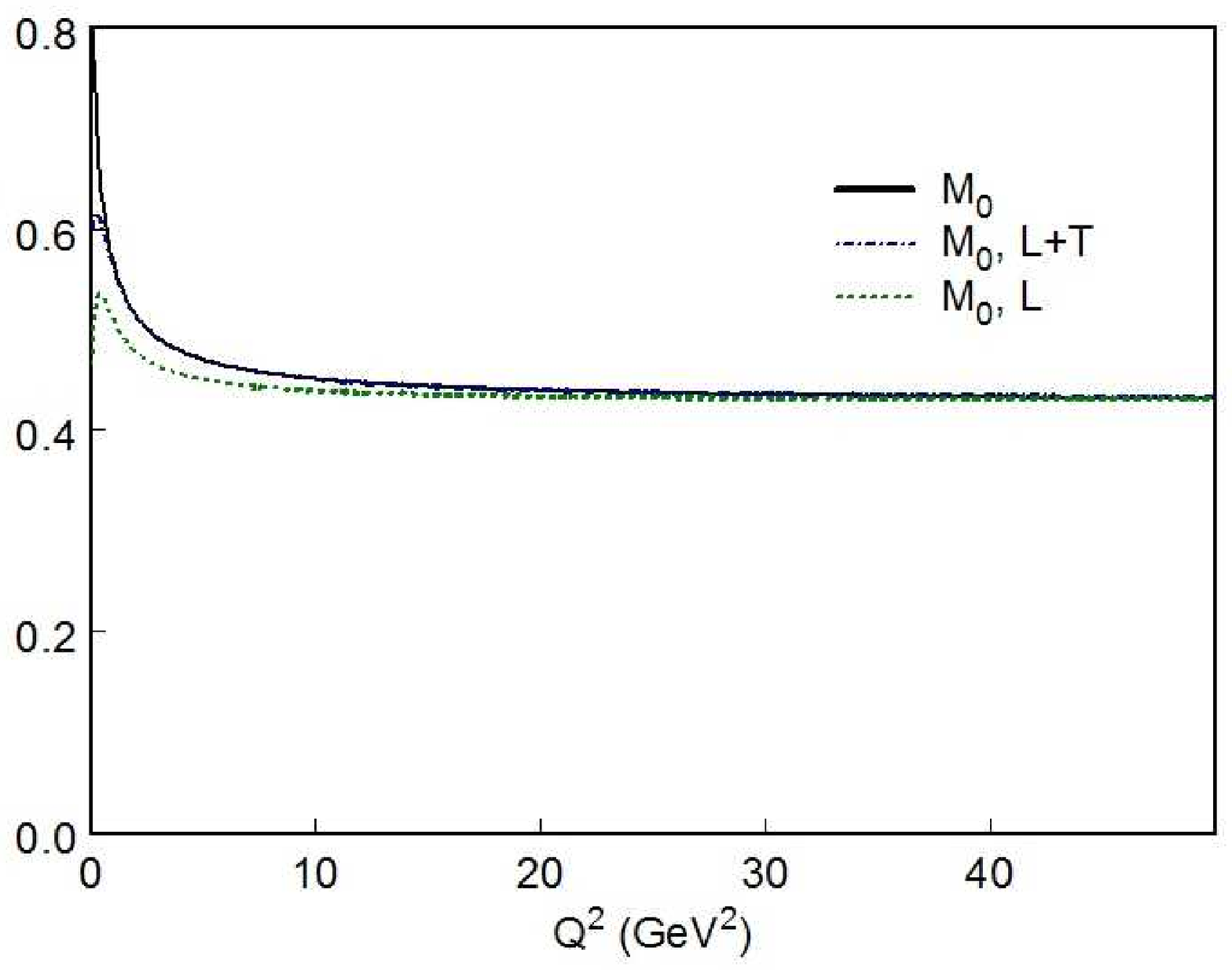}}
\epsfxsize=3.75in\centerline{\epsffile{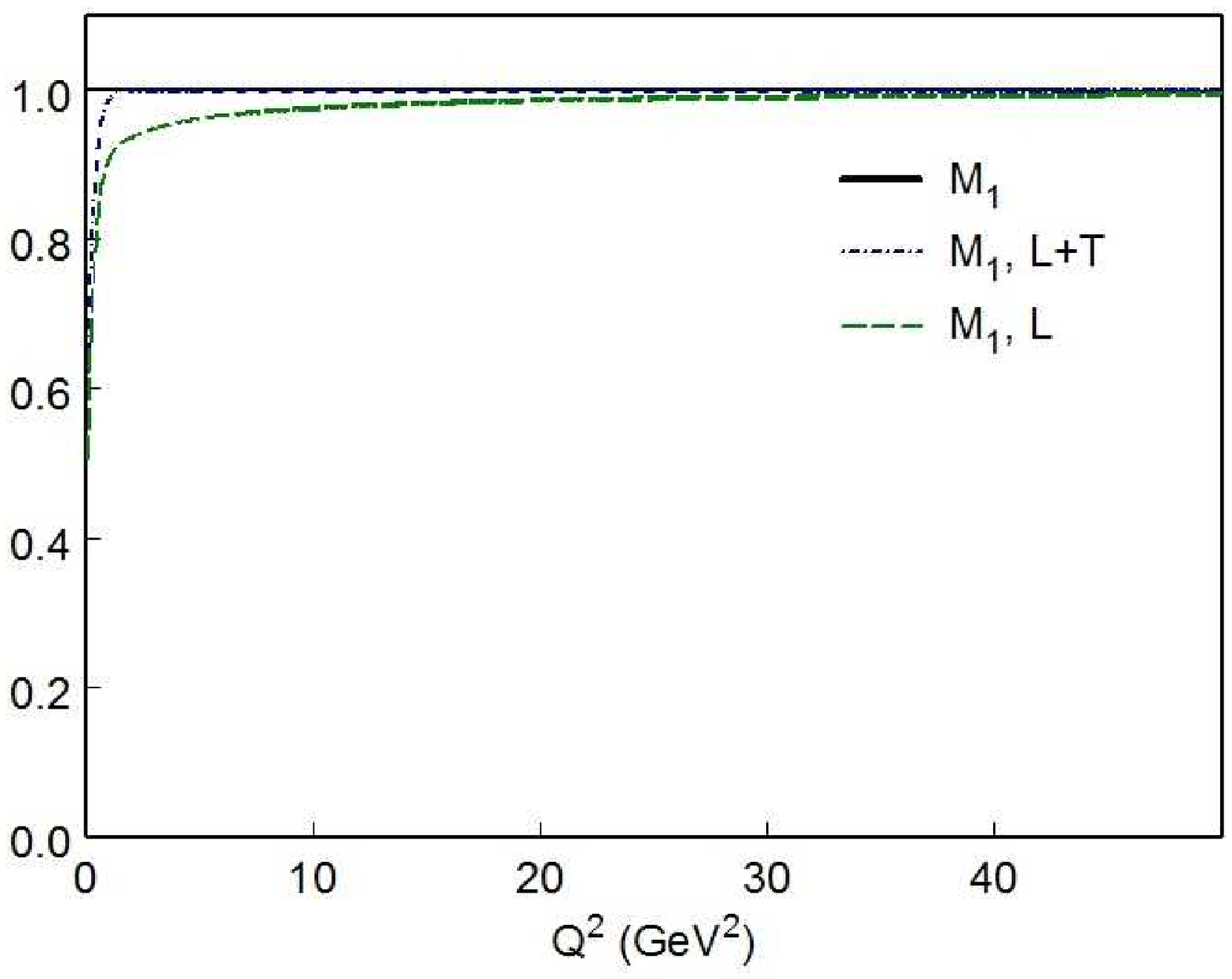}}
\epsfxsize=3.75in\centerline{\epsffile{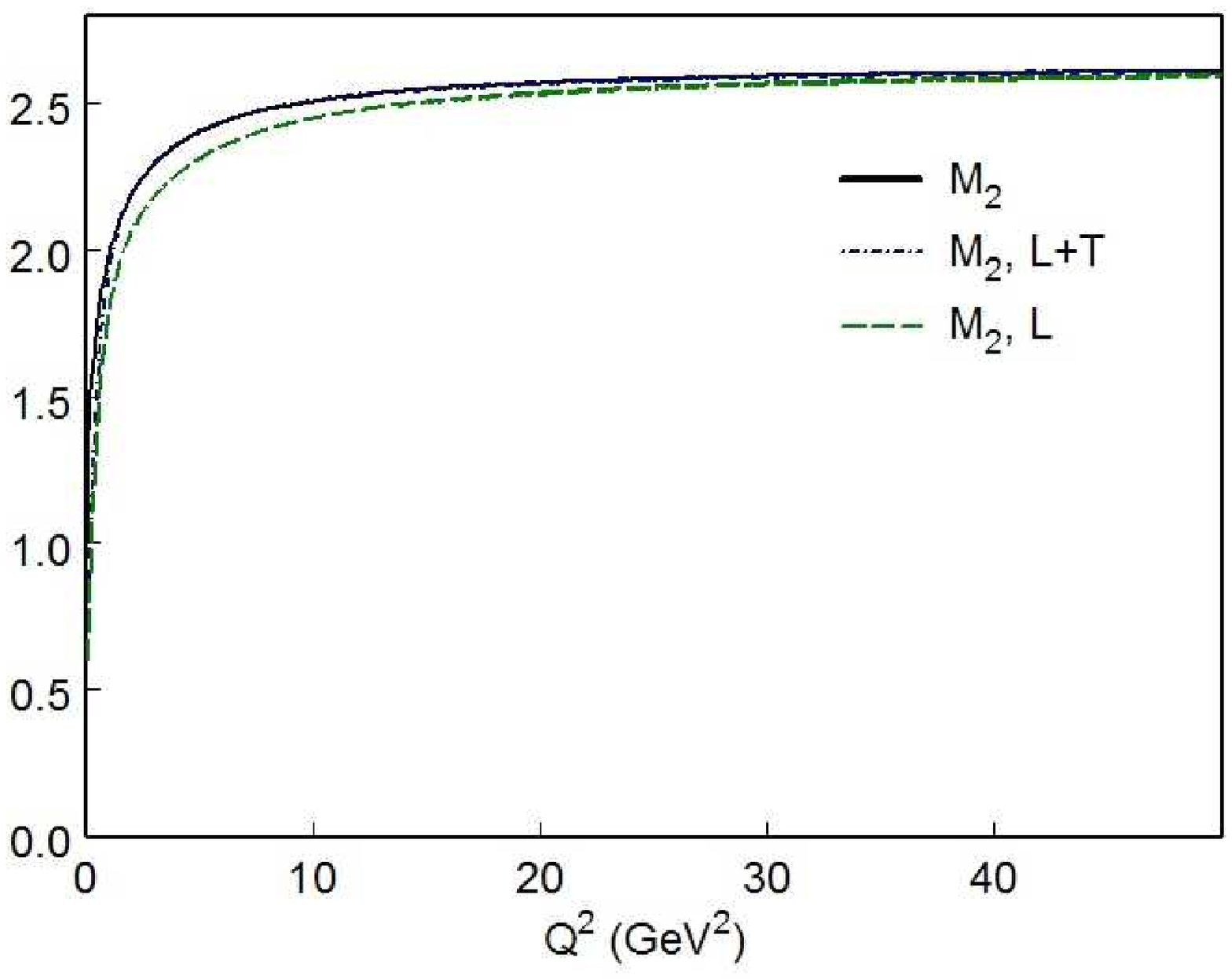}}
\caption{
The first three moments, $M_0$, $M_1$, and $M_2$. The solid line shows
the exact result calculated according to Eq. {\protect {\ref{moments}}}.
The dash-dotted lines show the moment calculated in the continuum
approximation, the dotted curves show the purely longitudinal contribution
to the moment in the continuum approximation.
Inclusion
of the transverse contributions improves convergence of the moments
to their asymptotic values.}\label{contmoments}
\end{figure}
In each panel the solid line represents the exact calculation
according to (\ref{moments}). The dash-dotted curve is a
calculation of the continuum approximation with both longitudinal
and transverse contributions, while the dashed curve includes only
the longitudinal contribution. Note that the continuum
approximation works very well down to a couple of GeV$^2$. Note
also that while the inclusion of the transverse contribution slows
convergence to the asymptotic value for $M_0$ it improves
convergence for the higher moments.

The continuum approximation can then be used to obtain an
expansion of the moments in powers of $1/Q^2$ reminiscent of the
operator product expansion (OPE) series,
\begin{equation}
M_n^{2k}(Q^2)=\sum_{i=0}^{k} \frac{c^{2i}_n}{Q^{2i}}\, .
\end{equation}
Note that we do not have any gluons in our model, and thus no
radiative corrections. The expansion coefficients $c^{2i}_n$
correspond to the non-perturbative matrix elements of higher
twist operators in the OPE.
Since this is an asymptotic series, the expansion will fail at low
$Q^2$ with the point at which the series diverges being dependent
upon the order of the series. The expansion coefficients for the
five lowest moments are shown in Table \ref{tabexpmom}.
Contributions to the coefficients from transverse and longitudinal
responses are shown along with the total.
\begin{table}
\caption{Leading coefficients of the expansion of the moments in
$1/Q^2$}\label{tabexpmom}
\begin{tabular}{llrrrr}
& & $c^0_n$ & $c^2_n$ & $c^4_n$ & $c^6_n$ \\ \tableline
$M_0$ & L &0.42859 & 0.20760 & -0.26022 & 0.13530\\
& T & 0.00000 & 0.13715 & -0.19201 & 0.19056\\
&total &0.42859 & 0.34475 & -0.45223 & 0.32586 \\ \tableline
$M_1$ &  L &1.00037 & -0.32052 &   0.95887 & -2.83879\\
&  T  &0.00000 &0.32012& -0.95870  &  2.83881\\
 &total & 1.00037&
-0.00039 &0.00017 &0.00002\\ \tableline
$M_2$ &L &  2.64117& -2.37461 &   5.53199 &-14.8231\\
&   T& 0.00000& 0.84518 &-3.22720 &10.6710\\
 & total& 2.64117& -1.52944& 2.30479&
 -4.1521\\\tableline
$M_3$ &L &7.6117 &-11.1960& 26.8417 &-71.9074\\
&   T& 0.0000 & 2.4357 &-11.2072& 40.7982\\
 &  total& 7.6117
&-8.7603 &15.6345& -31.1092\\\tableline
$M_4$& L &  23.5214 &-47.6636 &122.363 &-336.920\\
& T& 0.0000& 7.5268 &-40.233 &159.245\\
 & total &23.5214
&-40.1368 &82.130 &-177.675
\end{tabular}
\end{table}
The obvious feature of these coefficients is that they are not in
general small nor do they show any obvious convergence. The reason
for this can be seen from Fig. \ref{momentexpansion}. Here the
exact result is shown as a solid line and is compared to the
expansion with from one to four terms for both $M_0$ and $M_4$.
\begin{figure}
\epsfxsize=4.5in
\centerline{\epsffile{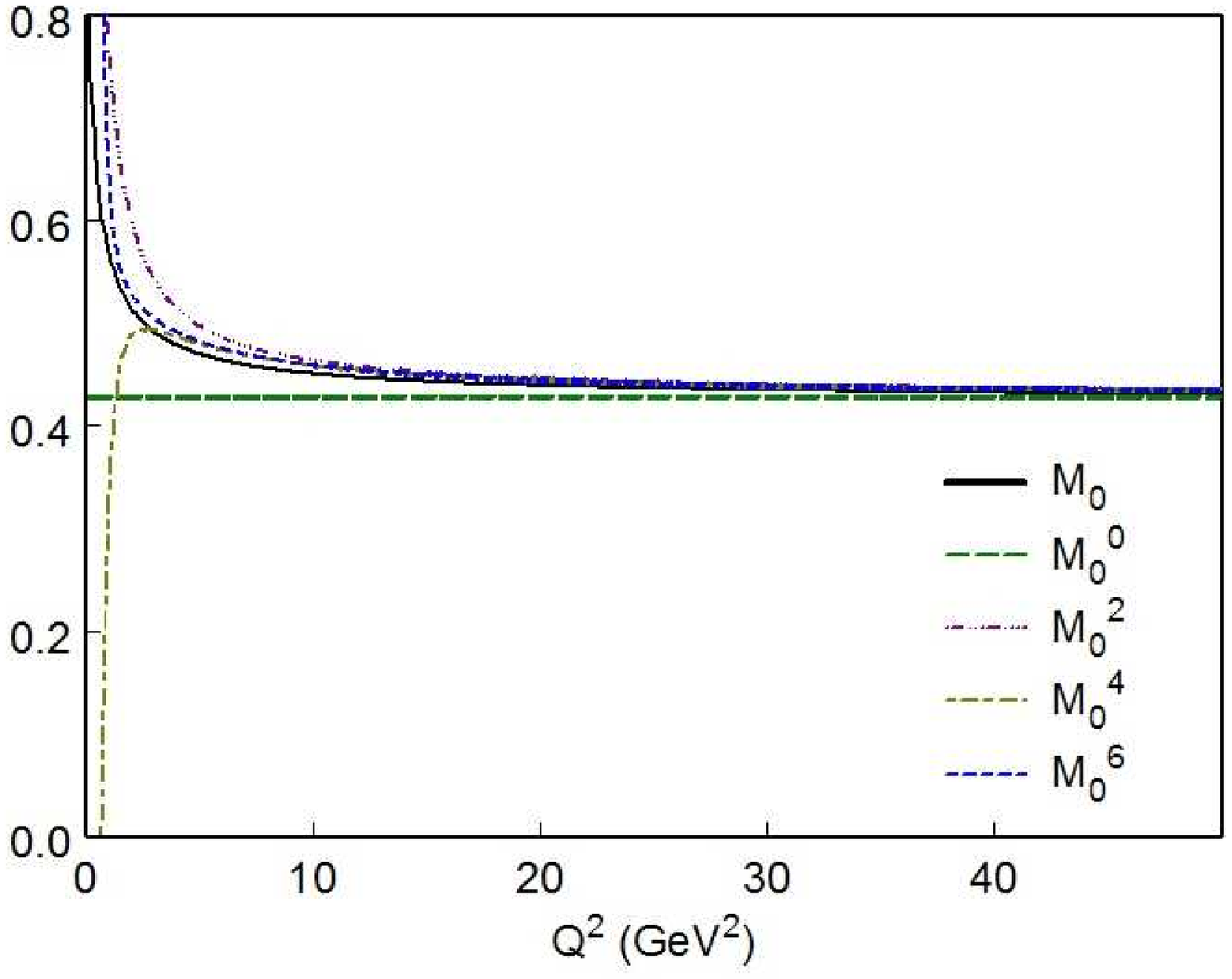}}\epsfxsize=4.5in\centerline{\epsffile{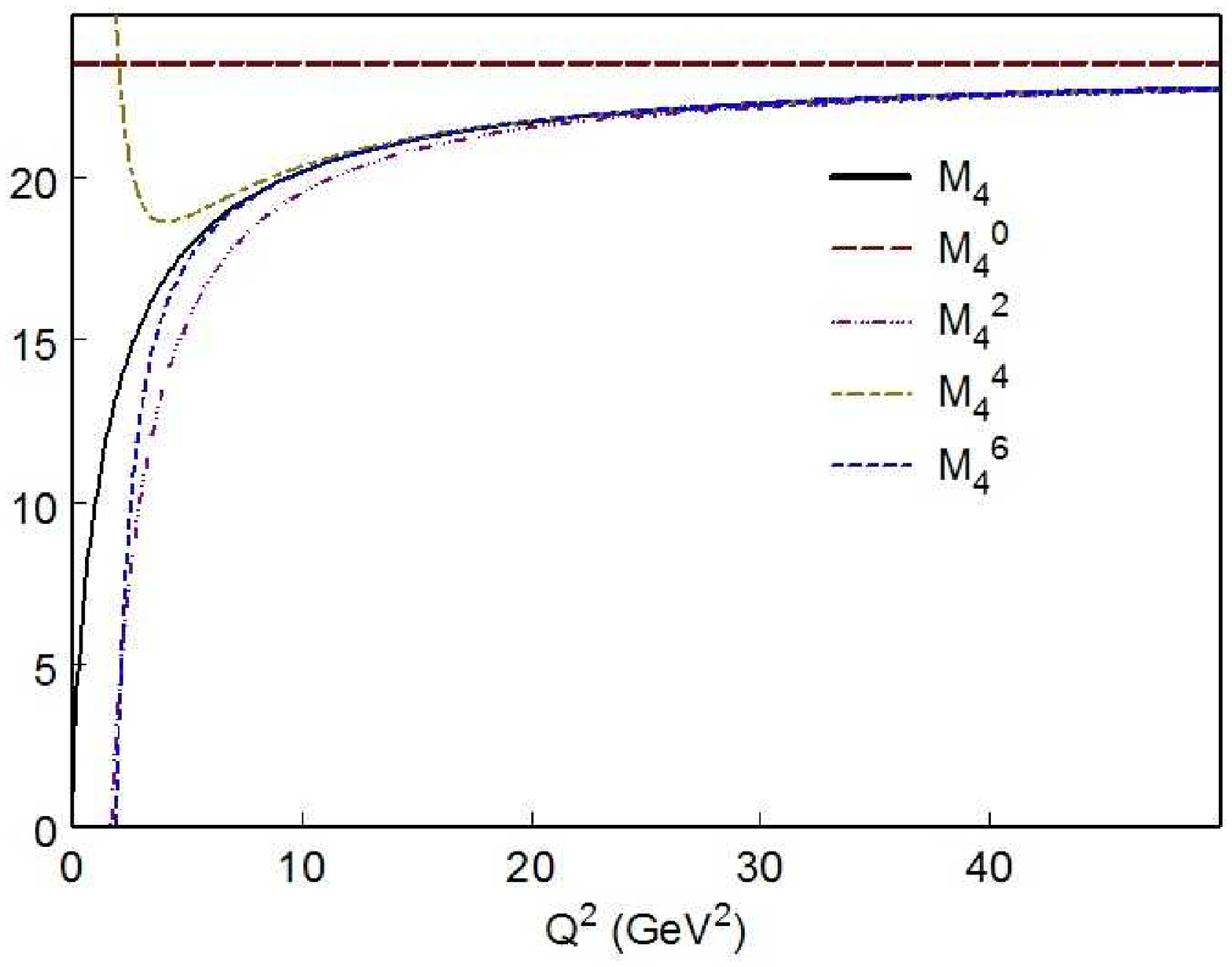}}
\caption{Comparison of moments to the expansion of the continuum
approximation in $1/Q^2$.}\label{momentexpansion}
\end{figure}
Since the moments all have a finite value at $Q^2=0$, the function
cannot be analytic in $1/Q^2$. Any expansion in this variable to a
finite number of terms will at some point diverge, either above or
below the correct result. Using an additional term to extend the
approximation to lower $Q^2$ must require that the coefficient of
this term be of opposite sign to the preceding term, leading to an
alternating series. Higher moments have the curvature toward the
finite result occurring at increasing values of $Q^2$. This
requires that the size of the coefficients for the higher terms in
the series must also be increasing. This shows that the global
duality observed in our model is the result of a delicate
cancellation between many ``higher twist'' terms. However, while
it is fascinating to speculate if duality in Nature is realized by
small higher order expansion coefficients or by cancellations, our
model is too simple to allow us to draw any conclusions about
this. In fact, results presented in \cite{jiunrau} indicate that
the expansion coefficients have the same sign. One may hope that
the building of more realistic models will allow us to gain a
better insight into this question in the future.

The relation asymptotic behavior of the moments to sum rules can
also be addressed in this model. Consider the moments of $F_2(u)$
\begin{equation}
{\cal M}_n \equiv\int_{-\infty}^\infty du \, u^{n-2} F_2(u) \,,
\end{equation}
where the integral starts at $-\infty$ to include contributions
from negative energy states as in the Coulomb Sum Rule. Using
(\ref{F2momdist}) this becomes
\begin{eqnarray}
{\cal M}_n  &=& \frac{m^2}{4\pi^2E_0} \int _{-\infty}^\infty du \,
u^n \int_{|E_0-mu|}^\infty dp \, p \, N(p)
=\frac{m^2}{4\pi^2E_0}\int_0^\infty dp \, p \,
N(p)\int_\frac{E_0-p}{m}^\frac{E_0+p}{m} du \, u^n\nonumber\\
&=& \frac{m^2}{4\pi^2E_0}\int_0^\infty dp \, p \, N(p)
\frac{1}{n+1}\left[\left(\frac{E_0+p}{m}\right)^{n+1}-
\left(\frac{E_0-p}{m}\right)^{n+1}\right]
\end{eqnarray}
The two lowest moments
\begin{equation}
{\cal M}_0=\frac{m}{2\pi^2 E_0}\int_0^\infty dp \, p^2 \,
N(p)=\frac{m}{E_0}=0.43002
\end{equation}
and
\begin{equation}
{\cal M}_1=\frac{1}{2\pi^2}\int_0^\infty dp \, p^2 \, N(p)=1
\end{equation}
are proportional to the normalization integral of the momentum
distribution.  Comparing these to the corresponding values of
$c^0_0$ and $c^0_1$ in Table \ref{tabexpmom} shows that the
contributions from negative energy states are small. Note also
that for a spin-1/2 constituent the expression corresponding to
(\ref{F2momdist}) have a leading factor of $u$ rather than $u^2$
as in this case. So the sum rules would be associated with $M_1$
and $M_2$ as expected.

\section{The low Q$^2$ Region}
\label{secloc}

After studying the scaling behavior of our model at high Q$^2$ and the
moments over a range of four-momentum transfers, we now discuss the
behavior at low Q$^2$. In this region, resonances are dominant for a
wide range in the scaling variable.

Before discussing the numerical results, a remark on the kinematics is
in order. For a fixed resonance in inclusive electron scattering from
the nucleon, its position in terms of Bjorken's scaling variable is
given by $ x_{res} = \frac{Q^2}{W_{res}^2 - M_N^2 + Q^2}$. This means
that for higher Q$^2$, the resonance position moves towards higher
values of $x_{Bj}$, and for very large Q$^2$, $x_{Bj} \rightarrow 1$.
In our case, the maximal value of the scaling variable $u$ is larger
than 1, and for very large Q$^2$, the resonances move out to very
large values of $u$, where their contribution is extremely small.

If local duality holds, we expect the resonance curve to oscillate
around the scaling curve and to average to it, once Q$^2$ is large
enough. In Bloom-Gilman duality, the finite energy sum rule gets
within 5 \% for Q$^2 \geq$ 1.75 GeV$^2$. For lower Q$^2$, the
resonances approach the scaling curve from below.  In our case, we
have the onset of scaling for larger values of Q$^2$ than observed in
Nature. This is not unexpected, as we consider an infinitely heavy
meson as target, and assume that this meson is made up of scalar
quarks. For this reason, our cross section for photon exchange for
large Q$^2$ is
dominated by the longitudinal part, and the transverse part,
comprising solely the convection current, is very small. In Nature,
for spin $1/2$ quarks, we have the magnetization current, which is
the dominant component of the cross section, and which therefore
determines the scaling behavior. So we cannot expect
our model calculation to show the same behavior as experimental
data for the same values of the four-momentum transfer.

\begin{figure}[!h,t]
\begin{center}
\epsfxsize=5in \centerline{\epsffile{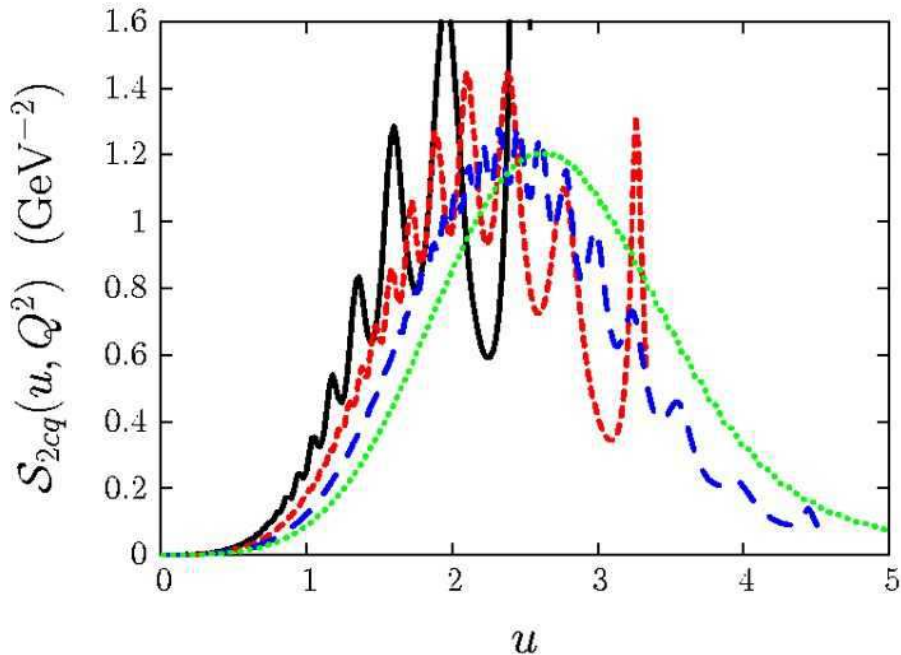}}
\end{center}
\caption{The low $Q^2$ behavior of ${\cal{S}}_{2cq} (u, Q^2)$ as a
function of $u$ for various values of $Q^2$. The solid  curve shows
$Q^2$ = 0.5 GeV$^2$, the short-dashed curve shows $Q^2$ = 1 GeV$^2$,
the long-dahsed  curve shows $Q^2$ = 2 GeV$^2$, and the dotted  curve
shows $Q^2$ = 50 GeV$^2$.}
\label{figlowq}
\end{figure}

In discussing local duality and resonances, the smoothing method used
becomes important. The visual appearance of ``resonances'' depends on
the chosen smoothing method: a bumpy structure is seen only when a
Breit-Wigner shape is inserted for the energy $\delta$-function.  It
also depends on the width chosen in the Breit-Wigner smoothing
method. For a smaller width, the resonances are visible for higher
Q$^2$. Depending on the width of the Breit-Wigner, e.g. for $\Gamma$ =
100 MeV, we do not see any resonances for Q$^2$ = 5 GeV$^2$, even
though this value of Q$^2$ is below the scaling region. In this paper,
the working definition of local duality which we use is ``resonance
curves oscillating around the scaling curve''. At some point, when
considering more realistic models, it may be useful and necessary to
introduce a sharper, more quantitative definition. However, at this
stage, we are interested more in qualitative results, and do not
intend to quantify how well exactly local duality works for our simple
model.

In Fig. \ref{figlowq}, we show our results for the scaling function
${\cal{S}}_{2cq} (u,Q^2)$ for various low values of $Q^2$.  The
$\delta$-function in the energy has been smoothed out using the
Breit-Wigner method, with a width of $\Gamma = 100$ MeV.  For visual
purposes, we have assigned a small width to the elastic peak, too.
One can see clearly from the figure that the resonances move out
towards higher $u$ with increasing Q$^2$, as dictated by
kinematics. While the elastic peak is rather prominent for Q$^2$ = 0.5
GeV$^2$ and Q$^2$ = 1.0 GeV$^2$, it becomes negligible for Q$^2 \geq$
2.0 GeV$^2$. This is the phenomenon we have observed already when
studying the moments: the elastic contribution there vanishes rapidly
with increasing Q$^2$.

\begin{figure}[!h,t]
\begin{center}
\epsfxsize=5in \centerline{\epsffile{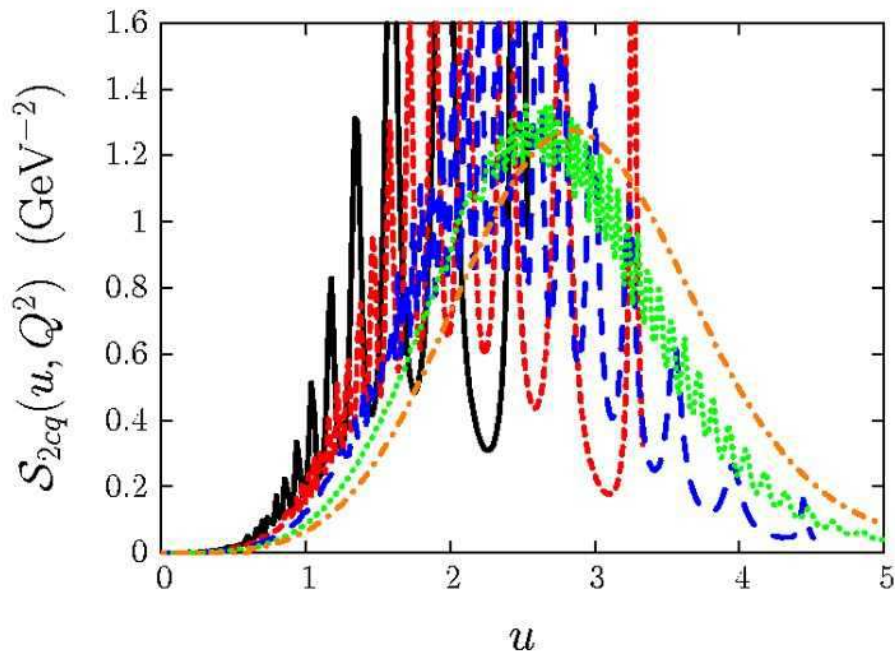}}
\end{center}
\caption{The low $Q^2$ behavior of ${\cal{S}}_{cq} (u, Q^2)$ as a
function of $u$ for various values of $Q^2$. The Breit-Wigner width
chosen to smooth out the energy $\delta$-functions is $\Gamma = 50
MeV$.  The solid  curve shows $Q^2$ = 0.5 GeV$^2$, the short-dashed
curve shows $Q^2$ = 1 GeV$^2$, the long-dashed curve shows $Q^2$ = 2
GeV$^2$, the dotted curve shows $Q^2$ = 5 GeV$^2$, and the dash-dotted
curve shows $Q^2$ = 50 GeV$^2$.}
\label{figlowqbw50}
\end{figure}

As already observed while studying the moments in the previous
section, the approach to scaling when using a virtual photon is slower
than for the all-scalar case discussed in \cite{dp1}. It is clear that
one needs to reach fairly large values of $Q^2$ before the ``resonance
curve'' averages with good accuracy to the scaling function. Indeed,
with our choice of Breit-Wigner width, this happens only when the
bumps have already disappeared, i.e. for $Q^2 \geq 5$ GeV$^2$.

In order to illustrate this point, we have included
Fig. \ref{figlowqbw50}, where we used a value of $\Gamma = 50$ MeV to
smooth out the energy $\delta$-function. The curves are more jagged
than for the larger width, and the $Q^2 = 5 $ GeV$^2$ curve still
shows plenty of resonance structure.

Overall, we find that the onset of local duality is definitely  slower
than for the all scalar case, which is what one expects due to the
additional structure in the more realistic case discussed in this
paper.

\section{Summary and Outlook}

We have presented a simple, quantum-mechanical model which allows us
to obtain the qualitative features of Bloom-Gilman duality. The model
assumptions we made are very basic: we assumed relativistic, confined
valence scalar quarks, and treated the hadrons in the infinitely
narrow resonance approximation. To simplify the situation further, we
did not consider a three quark ``nucleon'' target, but a target
composed of an infinitely heavy antiquark and a light quark. In
contrast to \cite{dp1} where all particles involved in the reaction -
electrons, photons and quarks - were considered to be scalar, we only
use scalar quarks in this paper. This makes the present model more
realistic - in particular, we were able to use a conserved current
here. However, there is still much work to be done in modelling.  The
goal of our model calculations is to gain qualitative insight into
duality, its applicability and accuracy in various kinematic regions,
not to quantitatively describe any data.
In the future, we plan to describe more realistic situations in our
model.  Note that our assumptions are very basic and general, so that
we will be able to extend our model in a straightforward manner.

There are several conditions that must be fulfilled in order to see duality.
In this paper, we put a special emphasis on three of these conditions:
we demonstrated how the transition from coherent scattering at low
Q$^2$ to incoherent scattering at high Q$^2$ takes place, we highlighted the
role of relativity by considering the contributions to the Coulomb sum rule
in a relativistic framework and a non-relativistic framework, and we gave
an analytic proof for the equality of the scaling curve in our model and
the parton model result.

Quark-hadron duality is not only very interesting in itself, it also
opens the door to very useful applications: duality relates the resonance
region data to data from the deep inelastic region. If duality is understood
well enough, and if the correct procedures for the averaging of the resonance
data and the attendant errors are established, we may exploit duality to gather
information in previously unreachable regimes. The investigation of
polarized structure functions in the high Bjorken-x region, $x_{Bj}
\rightarrow 1$, is a major part of the experimental program at Jefferson
Lab \cite{currentexps,12gevwp}. Even without knowing details about the
correct averaging procedures, it is clear from the experimental results
and our investigation of duality that the conventional, sharp distinction
between the ``resonance region'', corresponding to an invariant mass
$W < 2$ GeV,  and the ``deep inelastic region'' where $W > 2$ GeV, is
entirely artificial.

While quark-hadron duality has been investigated by theorists before,
modelling duality is an important new step in our way to a thorough
understanding of this phenomenon.
In the literature, one often finds the phrase that duality has been
explained in terms of QCD by DeRujula, Georgi, and Politzer \cite{dgp}.
What was
stated in their paper is that at moderate $Q^2$, the higher twist
corrections to the lower moments of the structure function are
small. The higher twist corrections arise due to initial and final
state interactions of the quarks and gluons.  Hence, the average value
of the structure function at moderate $Q^2$ is not very different from
its value in the scaling region.  While all this is true, the
statement is merely a rephrasing of the experimentally observed fact
that the resonance curve averages to the scaling curve in terms of the
language of the operator product expansion (OPE). However, the operator
product expansion does not explain why a certain correction is small or
why there are cancellations - the expansion coefficients which
determine this behavior are not predicted in the OPE. The ultimate
answer to this question might come from a numerical solution of QCD on
the lattice, but an understanding of the physical mechanism that leads
to the small values of the expansion coefficient in the framework of a
model is highly desirable. Also, the OPE will break down for very low
$Q^2$. Duality was experimentally observed \cite{jlab} to hold for $Q^2$
as low as 0.5 GeV$^2$ - a region where the validity of the OPE is
questionable. In our analysis of the moments and their expansion
coefficients, it became clear that a rigid application of the
OPE at very low Q$^2$ will inevitably lead to large, alternating
expansion coefficients.

The constant resonance to background ratio aspect of duality was
addressed in several papers by Carlson and Mukhopadhyay
\cite{carl}. They used counting rules to find the $Q^2$ dependence of
the form factors of the resonances in the Breit frame, and compared
them to the behavior of the scaling curve for large $x_{Bj}$ and to
the behavior of the background in the same region. From these
considerations, they could explain the constant ratio, provided the
$Q^2$-independent coefficients of the helicity amplitudes were not
anomalously small, as in the case of the $\Delta$ resonance, for which
the ratio vanishes. Still, there is no explanation why the coefficient
is small in one case and not in others, and there exist several models
with contradictory predictions.

The preceding observations clearly show the need for modelling. Even
though one may obtain expansion coefficients from calculations on the
lattice, an understanding of the underlying physical mechanisms will
most likely be gained only by considering models like ours. One great
advantage of a purely analytical model like the one presented here is
that explicit derivations of key quantities like the scaling function
are feasible.  The proof that the scaling function obtained for the
transition from a bound quark to an excited bound quark is the same as
the scaling function for the transition from a bound quark to a free
quark was given here for a linear potential, which is the relativistic
analog of a harmonic oscillator.  It is desirable to extend the
investigation to other types of potential, and to find a proof that
applies to a general class of potentials.  In a recent publication,
\cite{pp}, numerical methods were applied to study the responses of a
massless quark, and a disagreement between results with and without
final state interactions were observed.  This is in contrast to our
findings, and it is important to understand the reasons for these
differences.

The experimental data at very low Q$^2$ still average to one single
curve independent of their Q$^2$ \cite{jlab}. However, this is not
duality  in the sense defined in this paper because this curve
differs from the scaling curve. To investigate this interesting observation,
one must go beyond the model presented here, which contains valence
quarks only,
and therefore must produce a
valence-like shape. However, introducing sea quarks and modeling the
decay of the excited resonances, along with the corresponding
non-resonant production mechanisms from sea quark pairs, might shed
considerable light on this issue.

\acknowledgments

This paper is dedicated to the memory of our collaborator Nathan Isgur.

The authors thank F. Close, R. Ent, R. J. Furnstahl, N. Isgur,
 C. Keppel, S. Liuti, I. Niculescu, and W. Melnitchouk for stimulating
 discussions.   We thank E. Braaten and R. J. Furnstahl for useful
 comments on the manuscript. This work was in part supported by funds
 provided by the U.S. Department of Energy (D.O.E.)  under cooperative
 research agreement \#DE-AC05-84ER40150.

\appendix
\section{Duality in the Excitation Form Factors}
\label{ffapp}

In this appendix, we proceed to study
duality in the excitation form factor $F_{0,N} (\vec q ^2)$.
While this duality is not directly related to an
observable like the structure functions or response functions, it
exhibits duality very clearly. Duality in the form factors
 has recently received some attention in
\cite{closeisgur}.

 The duality prediction is that it should not
matter if we describe the process in question in a
perturbative QCD picture, involving only a free quark,
or in a hadronic picture with resonances.
The form factor for a hypothetical free quark is just 1, as it does
not have any structure. In the hadronic picture, we have inclusive
electron scattering where we can excite all resonances -- as the final
hadronic state is not observed, we have to sum over the resonances
incoherently. So we have to compare $\sum_N |F_{0,N} (\vec q \, ^2)|^2
$ to 1. Fig. \ref{figsingleresff} shows single form factors for the
lowest resonances, the elastic peak and inelastic excitations up to N
= 5. All form factors look qualitatively similar, except for the N=0
elastic form factor, which starts at 1 for $|\vec q| = 0$.  In
general, the form factors increase in width and decrease in height
with increasing N.

\begin{figure}[!h,t]
\begin{center}
\leavevmode
\epsfxsize=5in \centerline{\epsffile{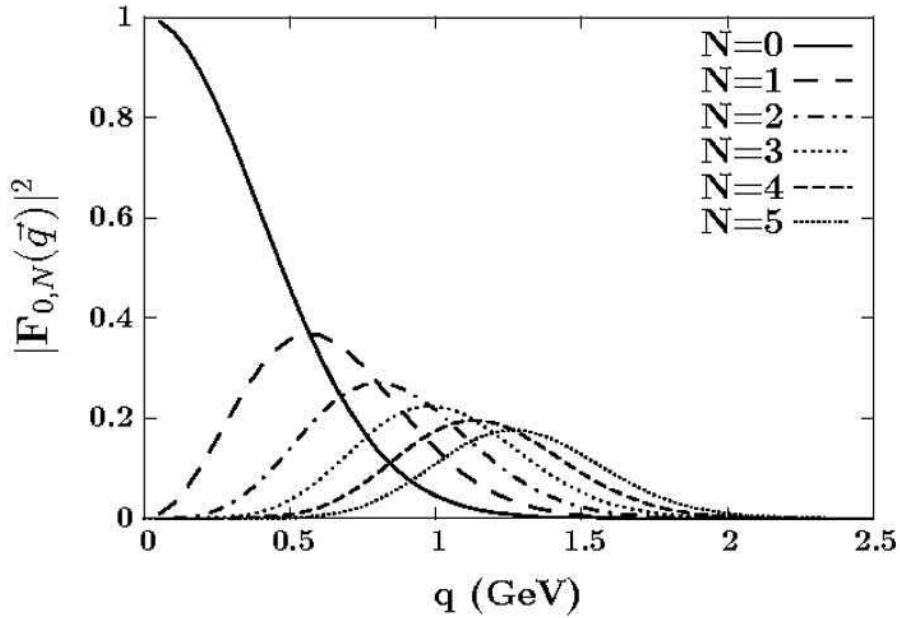}}
\end{center}
\caption{The excitation form factor squared, $ |F_{0N} ( \vec q \,
)|^2$ for the lowest excitations, N~=~0 to N~=~5, and $\beta = 0.4 $ GeV.}
\label{figsingleresff}
\end{figure}

Using our previous expression for the form factor, eq. (\ref{eqff}), we
find for the sum up to a certain value $N_{max}$
\begin{equation}
\sum_{N=0}^{N_{max}} |F_{0N} ( \vec q \, )|^2  =
 \exp{(-\frac{\vec q \, ^2}{2 \beta^2})}
\sum_{N=0}^{N_{max}}  \frac{1}{N!}  \left (\frac{\vec q \, ^2 }
{2 \beta^2} \right )^N \, ,
\end{equation}
and it is obvious that
\begin{equation}
\sum_{N=0}^{N_{max}} |F_{0N} ( \vec q \, )|^2  = 1 \, \, \mbox{if}
 \, \, \, N_{max} \rightarrow \infty \,,
\end{equation}
as mentioned when discussing the Coulomb sum rule.
However, we are limited in the maximal value of $N_{max}$ not by
technical problems, but by a physical constraint: we are considering
electron scattering, i.e. space-like kinematics, and therefore
we must fulfill the condition
\begin{equation}
  Q^2 > 0 \Leftrightarrow |\vec q| > \nu = E_N - E_0 \, ,
\end{equation}
so that for fixed three-momentum transfer $|\vec q|$, we find a limit
on the value of N. The form factor sums are shown in
Fig. \ref{figffsumrel} for various values of the oscillator parameter
$\beta$. Larger values of $\beta$ indicate a stronger binding.

\begin{figure}[!h,t]
\begin{center}
\epsfxsize=5in \centerline{\epsffile{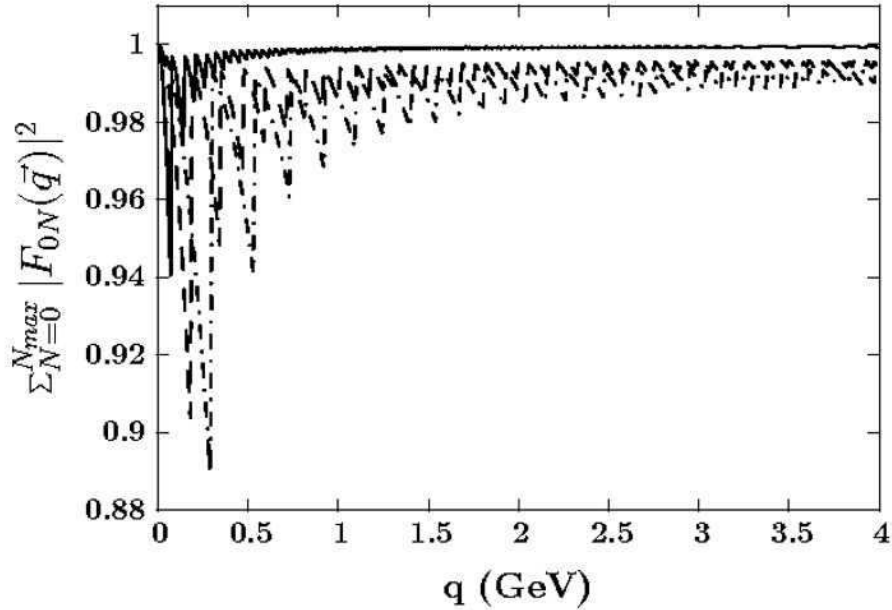}}
\end{center}
\caption{The sum over the excitation form factors squared,
for $\beta = 0.2 $ GeV (solid line),  $\beta~=~0.4$~GeV (dashed line),
and $\beta = 0.6 $ GeV (dash-dotted line).}
\label{figffsumrel}
\end{figure}

The spiky character of the curves stems from the fact that the form
factors for the N-th state are only allowed to contribute to the sum
if $|\vec q| > \nu_N$, where $\nu_N := E_N - E_0$. Therefore, the sum
jumps up whenever another threshold is crossed. This effect can be
observed best for the strongest binding, $\beta = 0.6 $ GeV, as the
gaps between the energy levels are largest in this case. With
increasing value of the three-momentum transfer, more and more
resonance states can be excited and contribute to the sum: the spikes
subside and the average value of the sum gets fairly close to 1. The
curve for the weakest binding, $\beta = 0.2 $ GeV, becomes almost
smooth and takes on a value of 0.9994, i.e. duality is violated by
less than 0.06 \%. For $\beta = 0.4 $ GeV, the sum reaches 0.995, so
duality is violated only by 0.5 \%. Even for the strongest binding, $\beta =
0.6 $ GeV, the duality prediction is fulfilled within 1\%. Here, we see
 a typical feature of duality, the need for many
resonances to contribute in order to reproduce the behavior of a free
quark. At low $|\vec q|$, where only a few resonances can contribute,
the deviations from 1 are larger.  The fact that duality is fulfilled
best for weak binding is what we expect: a quark that is bound very
lightly and then receives a hard momentum transfer behaves essentially
as if it was free. If the binding gets stronger, the situation gets more
non-perturbative, and duality does not work as well.

The duality as seen in the form factors is reminiscent of duality in the
decay rates of the semileptonic decay of heavy quarks
\cite{isgurwise,isgurplb}. There, in the
limit of infinite masses of the $b$ and $c$ quarks, the loss of strength
in the elastic channel is compensated for by the increase in the inelastic
decay channels. Once one considers heavy, but not infinitely heavy, quark
 masses, one obtains a jagged structure, with peaks getting close to the
free quark limit, quite similar to what we observe when
considering the Coulomb Sum Rule
and the excitation form factors.

\begin{figure}[!h,t]
\begin{center}
\epsfxsize=5in \centerline{\epsffile{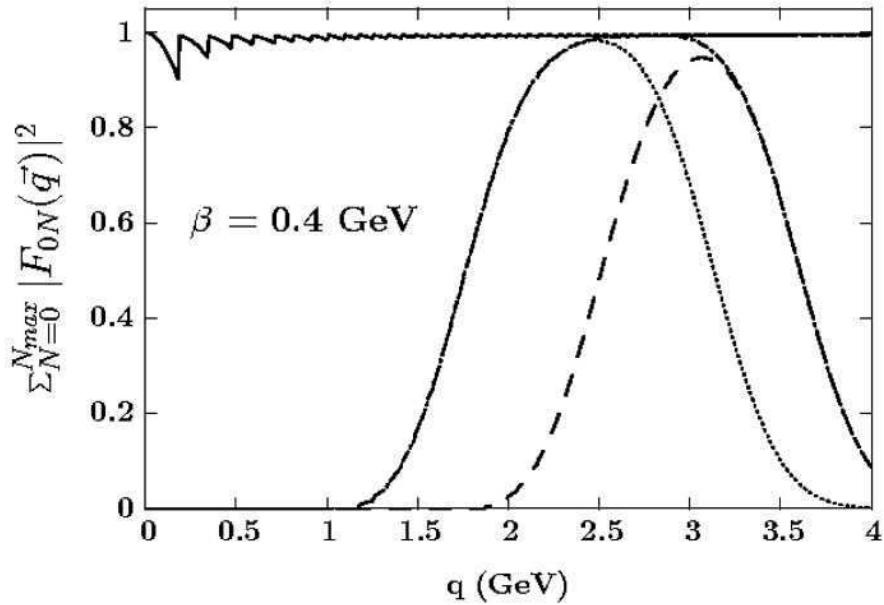}}
\end{center}
\caption{The sum over the excitation form factors squared $|F_{0,N}
(\vec q ^2)|^2$
for $\beta = 0.4 $ GeV. The solid line shows the sum over all allowed N,
the dash-dotted line shows the sum from N = 10 to N = 40, the dotted line
shows the sum from N = 10 to N = 30, and the dashed line shows the
sum from  N = 20 to N = 40.}
\label{figffsumtrunc}
\end{figure}

Let us consider the number of resonances needed in more detail, so
that we can draw further conclusions on the kind of duality we are
observing here. In the calculations presented in
Fig. \ref{figffsumrel}, we summed up to the highest allowed N,
N$_{max}$, which is quite large in general. In
Fig. \ref{figffsumtrunc}, we present the full curve, and three curves
where we summed over a limited number of resonances only, namely from
N = 10 to N = 40 (dash-dotted line), N = 10 to N = 30 (dotted line)
and N = 20 to N = 40 (dashed line). One can see that for a small
interval in three-momentum transfer of $\vec q = 2.6$ GeV - $ 3.0 $ GeV,
it is sufficient to include only resonances from N = 10 to N = 40 in
order to reproduce the full, unrestricted curve. One also sees from the
dotted line that for lower three-momentum transfer, the inclusion of
just the twenty resonances from N = 10 to N = 30 suffices to get close
to the full curve $\vec q = 2.6 $ GeV, while the same number of
resonances does not suffice to approximate the full curve at a
slightly higher value of $|\vec q|$ (see the dashed line).

So, while it is clear that the ``degrees of freedom duality'' holds
very nicely over the whole kinematic range, we see that duality - the
truncated version - does not hold as well: we do need a certain number
of resonances to obtain duality in a limited kinematic interval, and
this number increases when we increase the three-momentum transfer.

\section{The Role of Relativity}

We have stressed the importance of relativity before, and while it is
quite obvious that one needs to include it for GeV momentum transfers
to light quarks, it is instructive to see how relativity works for the
form factors. In order to illustrate this point, we show the sum of
the excitation form factors squared calculated for the
non-relativistic harmonic oscillator potential in
Fig. \ref{figffsumnr}. As mentioned in section \ref{secmod}, the wave
functions, and therefore the form factors, are the same. The
difference lies in the energy spectrum.

\begin{figure}[!h,t]
\begin{center}
\epsfxsize=5in \centerline{\epsffile{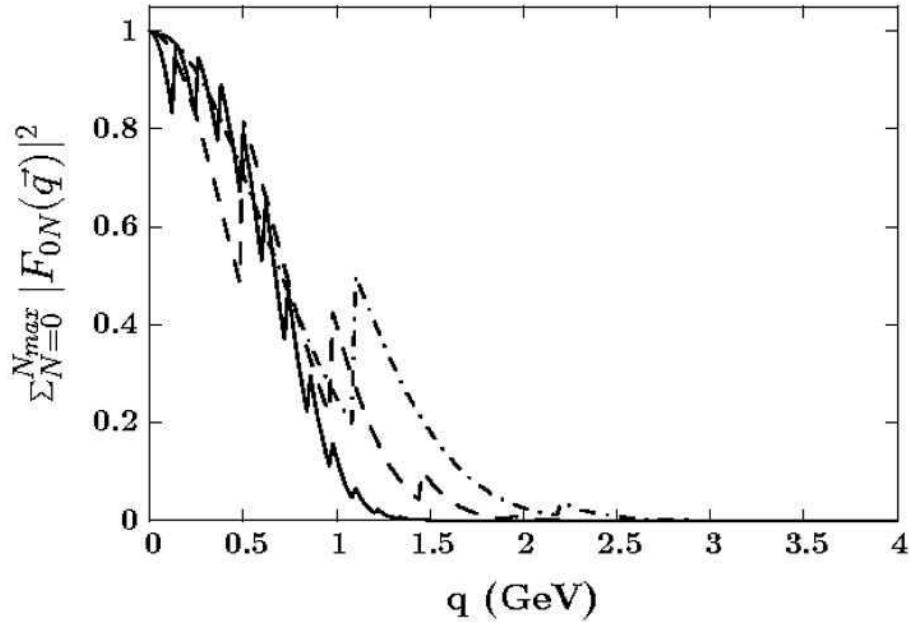}}
\end{center}
\caption{The sum over the excitation form factors squared for the
non-relativistic case,
for $\beta = 0.2 $ GeV (solid line),  $\beta = 0.4 $ GeV (dashed line),
and $\beta = 0.6 $ GeV (dash-dotted line). Note that the scale differs
from the scale in \protect{Fig. \ref{figffsumrel}}.}
\label{figffsumnr}
\end{figure}

Obviously, duality in the non-relativistic case does not work at all:
the curves start out at 1, as the elastic form factor for $|\vec q| =
0$ is 1, but then fall off immediately. Whenever a new threshold
opens, the additional contribution is not sufficient to compensate the
fall-off of the other form factors: they do not contribute at their
maximum value, but only with the ``high $|\vec q|$'' side of the peak,
where the form factor drops quickly, see Fig. \ref{figsingleresff}. In
the non-relativistic case, the spacing between the energy levels is
wider than in the relativistic case, where the levels shrink together.
This means that considerably fewer resonances are allowed to
contribute at the same, fixed $|\vec q|$, and therefore the resulting
sum is much smaller and duality is violated.  To clarify this point,
we show a comparison of energy levels in Fig.~\ref{figenercomp}.

\begin{figure}[!h,t]
\begin{center}
\epsfxsize=7in \centerline{\epsffile{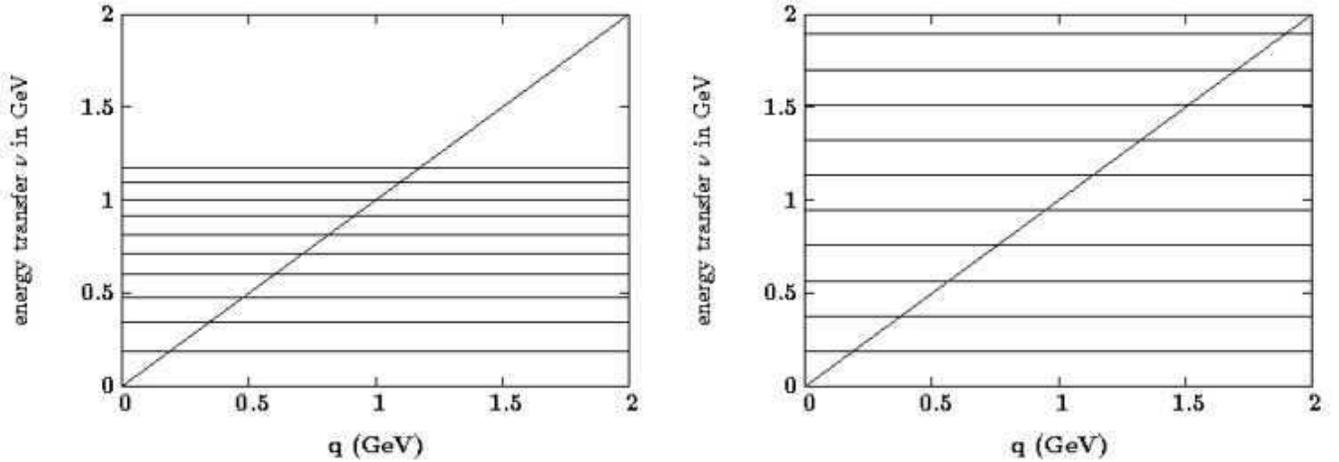}}
\end{center}
\caption{The first eleven energy transfers $\nu$ for the relativistic
case with $\beta = 0.4 $ GeV (left panel) and for the non-relativistic
case with $\beta = 0.25 $ GeV (right panel). The diagonal line in both panels
indicates the photopoint, i.e. $|\vec q| = \nu$. All energy transfers below
this line are allowed in the space-like region.}
\label{figenercomp}
\end{figure}

In the left panel, we show the energy transfers $\nu$ for the first
eleven energy levels (elastic and the first ten resonances) of a
relativistic harmonic oscillator with $\beta = 0.4 $ GeV.  The diagonal
line marks the photopoint, $|\vec q| = \nu$.  This means that for a
given $|\vec q|$, all the energy transfers below that line are in the
space-like region and therefore allowed.  E.g., in the relativistic
case for $|\vec q| = 1 $ GeV, nine resonances (counting the elastic)
can be excited. The right panel of Fig. \ref{figenercomp} shows the
energy transfers $\nu$ for the energy levels of a non-relativistic
harmonic oscillator with $\beta = 0.25 $ GeV. We chose the $\beta$ for
the non-relativistic oscillator in order to reproduce the energy
splitting between the ground state and the first excited state of the
relativistic case. One clearly sees that fewer resonances can
contribute here, e.g. only six compared to nine in the relativistic
case at $|\vec q| = 1 $ GeV. The discrepancy grows larger for higher
$|\vec q|$, as the relativistic energy levels move closer together,
while the non-relativistic ones are equally spaced.

In conclusion, we have seen that the relativistic description is
necessary to ensure a correct treatment of the phase space. Only with
a proper relativistic phase space do we see duality in the excitation
form factors. This was already clear from our discussion or the
Coulomb Sum Rule in the main body of the paper.

Mathematically, degrees of freedom duality in the excitation form
factor means that one can expand a plane wave (free quark) in a set of
Hermite polynomials (bound quark), provided one uses a sufficiently
large number of basis states. Any other set of ortho-normal
polynomials would also work.

\end{document}